\documentclass[epj]{svjour}
\usepackage{amsmath}
\usepackage{amssymb}
\usepackage{cite}
\usepackage[english]{babel}
\usepackage{float}
\usepackage{dcolumn}

\usepackage{slashed}

\usepackage{epsfig}

\def\lsim{\mathrel{\rlap{\lower4pt\hbox{\hskip1pt$\sim$}}
    \raise1pt\hbox{$<$}}}                % less than or approx. symbol
\def\gsim{\mathrel{\rlap{\lower4pt\hbox{\hskip1pt$\sim$}}
    \raise1pt\hbox{$>$}}}                % greater than or approx. symbol
\begin{document}

\title{Thermo-magnetic effects in quark matter: Nambu--Jona-Lasinio model constrained
by lattice QCD} 

\author{Ricardo L.S. Farias\inst{1,2}, Varese S. Tim\'oteo\inst{3}, Sidney S. Avancini\inst{4}, 
Marcus B. Pinto\inst{4}, and Gast\~ao Krein \inst{5}}

\institute{Departamento de F\'isica, Universidade Federal de Santa Maria, 
97105-900 Santa Maria, RS, Brazil
\and Physics Department, Kent State University, Kent, OH 44242, USA  
 \and Grupo de \'Optica e Modelagem Num\'erica - GOMNI, Faculdade de Tecnologia - FT, \\
Universidade Estadual de Campinas - UNICAMP, 13484-332 Limeira, SP , Brazil  
\and Departamento de F\'{\i}sica, Universidade Federal de Santa
  Catarina, 88040-900 Florian\'{o}polis, Santa Catarina, Brazil  
\and Instituto de F\'{\i}sica Te\'orica, Universidade Estadual
  Paulista,  Rua Dr. Bento Teobaldo Ferraz, 271 - Bloco II \\ 
  01140-070 S\~ao Paulo, SP, Brazil}
\date{Received: date / Revised version: date}

\abstract{
The phenomenon of inverse magnetic catalysis of chiral symmetry in 
QCD predicted by lattice simulations can be reproduced within the 
Nambu--Jona-Lasinio model if the coupling~$G$ of the model decreases 
with the strength $B$ of the magnetic field and temperature~$T$. The 
thermo-magnetic dependence of $G(B,T)$ is obtained by fitting recent 
lattice QCD predictions for the chiral transition order parameter. Different 
thermodynamic quantities of magnetized quark matter evaluated with 
$G(B, T)$ are compared with the ones obtained at constant coupling, $G$. 
The model with  $G(B,T)$ predicts a more dramatic chiral transition as 
the field intensity increases. In addition, the pressure and magnetization 
always increase with $B$ for a given temperature. Being parametrized by 
four magnetic field dependent coefficients and having a rather simple exponential 
thermal dependence our accurate ansatz for the coupling constant can be easily 
implemented to improve typical model applications to magnetized quark matter.
}


\authorrunning{R.L.S. Farias et al.}
\titlerunning{Thermo-magnetic effects in quark matter}
\maketitle

The fact that strong magnetic fields may be generated in peripheral  heavy-ion 
collisions~\cite{heavyion1,heavyion2} and may also  be present in magnetars~\cite {magnetars} 
has motivated many recent investigations regarding the effects of a magnetic field in defining  
the boundaries of the quantum chromodynamics (QCD) phase diagram {\textemdash} for recent reviews, 
see Refs.~\cite{{Miransky:2015ava},{revandersen}}. In both situations, the  magnitude of the 
magnetic fields is huge and may reach respectively $10^{19} \,{\rm G}$ and $10^{18} \,{\rm G}$. 
In heavy ion collisions, the presence of a strong magnetic field most certainly plays a role, 
despite the fact that the field intensity might decrease very rapidly, lasting for about 
$1-2~{\rm fm}/{\rm c}$ only~\cite{heavyion1,heavyion2}. 
The possibility that this short time interval may~\cite{tuchin} or may not~\cite{mclerran} be
affected by conductivity remains under dispute.  Very interesting effects are also expected in neutron
stars with a possible quark core, as the magnetic field penetrates into the quark core
in the form of quark vortices due to the presence of Meissner currents \cite{david1,david2}. 
At~zero temperature, the great majority of effective models for QCD are in agreement with respect 
to the occurrence of the phenomenon of magnetic catalysis (MC), which refers to the increase of 
the chiral order parameter represented by the (light) quark condensates with the strenght $B$ of 
the magnetic field. On the other hand, at finite temperature such models fail to predict the 
inverse magnetic catalysis (IMC), an effect discovered by lattice QCD (LQCD) 
simulations~\cite{lattice,fodor}, in that the pseudo-critical temperature $T_{pc}$ for chiral 
symmetry partial restoration decreases as $B$ increases. We remark that LQCD simulations as well as  
most models predict that the crossover transition observed at $B=0$ and vanishing chemical potential 
survives the presence of a background magnetic field, at least for realistic field intensities. The 
failure of model calculations in predicting IMC at high temperatures
has motivated a large body of work attempting to clarify the reasons for the observed 
discrepancies 
\cite{Miransky:2015ava,revandersen,
fragaMIT,prl,kojo,endrodi,leticiaLN,EPNJL1,EPNJL2,EPNJL3,debora,bruno,bonati,ayala1,
ferreira1,ayala2,ayala3,ferrer1,cao1,sadooghi1,andersen2,kanazawa,ayala4,endrodi2,huang1,
feng1,braum1,Pawlowski1,ayala5,endrodi3,debora2,noronha1,marcus1,debora1,debora3,ayala6,
Cao2,Yoshiike,Hattori1,feng2,cao3,fengli1,raya}. 
Intuitively, it is natural to attribute
the failure to the fact that most effective models lack gluonic degrees of freedom and so are 
unable to account for the back reaction of sea quarks to the external magnetic field. This implies,
in particular, the absence of asymptotic freedom, a key feature of QCD that plays an
important role in processes involving high temperatures and large baryon densities, and, of 
course, large magnetic fields. Since long ago, such effects have been mimicked in effective models 
by making the coupling strength to decrease with the temperature and/or 
density according to some ansatz~\cite{bernard,marcus}. 
More recently, this very same strategy was adopted in the case of hot magnetized quark 
matter. In particular, in Ref.~\cite{PRCfarias2014}, the IMC phenomenon found by lattice 
simulations was explained within the two-flavor Nambu--Jona-Lasinio model (NJL) when the 
coupling constant, $G$, is forced to decrease with {\it both} the magnetic field strength $B$ 
and the temperature $T$, simulating effects not captured with the conventional NJL model.
A similar procedure was used with a $SU(3)$ Polyakov-NJL (PNJL) model, but with $G$ depending only 
on the magnetic field~\cite{prdferreira}; this leads, however, to a non monotonic decrease of $T_{pc}$ 
at high field values. In a very recent work~\cite{ayala7}, an explicit calculation of the one-loop 
correction to the quark-gluon vertex has shown that competing effects between quark and gluon color 
charges make the effective quark-gluon coupling to decrease as the strength of the magnetic field 
increases at finite temperatures. This certainly lends strong support to the idea~\cite{PRCfarias2014} 
that the IMC is  due to the decrease of the effective coupling between quarks and gluons in the presence 
of magnetic fields at high temperatures.

In the present paper we investigate the implications of using a $B-$ and $T-$modified NJL coupling 
for thermodynamic quantities of magnetized quark matter. We are particularly interested in 
the qualitative changes that a $G(B,T)$ causes in quantities very sensitive to the chiral
transition, such as the speed of sound, thermal susceptibility and specific heat. This is an
important open question since the interaction that is implied by a $G(B,T)$ gives rise to a 
new phenomenology that has not been fully explored in the literature. The investigation of the
correlation between a $T$ and $B$ dependence of the NJL coupling $G$ used to describe IMC with 
other physical quantities is important to get further insight into the role played by effects not 
captured by the normal NJL. 
As we shall show, the very same $G(B,T)$ required to fit the lattice result for $T_{pc}$, gives 
results for the pressure, entropy and energy density that are in qualitative agreement with 
corresponding lattice results, while a $B-$ and $T-$independent coupling gives qualitatively 
different results for those quantities. This seems to be a clear indication that the $B$ and $T$ 
dependence in $G$ needed to describe $T_{pc}$ is neither fortuitous nor valid for a single 
physical quantity only; it seems to capture correctly the physics left out in the conventional 
NJL model. Instead of the parametrization used in Ref.~\cite {PRCfarias2014}, based on qualitative 
arguments referring to asymptotic freedom, in the present paper we base the parametrization of~$G$ 
on a precise fit of recent LQCD calculations. In doing so, one avoids any particular interpretation 
on the effects behind fitting formulas used for the $B$ and $T$ dependence of~$G$, as any interpolation
formula of the lattice data points leads to qualitatively similar results for the thermodynamical 
quantities. We fit LQCD results for the magnetized quark condensates with a particularly simple 
Fermi-type distribution formula for $G(B,T)$, parametrized by four $B$-dependent coefficients. As 
we shall demonstrate, one of the main physical implications of using such thermo magnetic effects in the  coupling constant is 
that the signatures associated with the chiral transition in thermodynamic quantities become more 
markedly defined as the field strength increases. Also, our results for the pressure and magnetization 
are in line with LQCD predictions, which find that at  a fixed temperature, these quantities always 
increase with $B$. This behavior, especially close to the transition region, is not observed with the NJL 
model with a $B$ and $T$ independent coupling~$G$. 

In the next section we review the results for the magnetized NJL pressure within the mean 
field approximation (MFA). In Sec.~III we extract  $G(B,T)$ from an accurate 
fit of LQCD results. Numerical results for different thermodynamical quantities are presented 
in Sec.~IV. Our conclusions and final remarks are presented in Sec.~V.

%%%%%%%%%%%%%%%%%%%%%%%%%%%%%%%%%%%%
\section{Magnetized NJL Pressure}
\label{njlTB}

Here we consider the isospin-symmetric two flavor  version of the NJL model~\cite{njl}, 
defined by the Lagrangian density
\begin{eqnarray}
\mathcal{L}_{\rm NJL} &=& - \frac{1}{4}F^{\mu\nu}F_{\mu\nu}  + {\bar \psi} \left( {\slashed D} -m\right) \psi 
\nonumber \\
&+& G\left[ (\bar \psi\psi)^{2}  + (\bar\psi i \gamma_5{\bf\tau}\psi)^{2}
\right],
\label{njl2}
\end{eqnarray}
\noindent
where the field $\psi$ represents a flavor iso-doublet of $u$ and $d$ quark flavors and 
$N_{c}$-plet of quark fields, ${\bf\tau}$ are the isospin Pauli matrices,
$D^\mu =(i\partial^{\mu}-QA^{\mu})$ the covariant derivative, Q=diag($q_u$= $2 e/3$, $q_d$=-$e/3$) 
the charge matrix and  $A^\mu$, $F^{\mu\nu} = \partial^\mu A^\nu - \partial^\nu A^\mu$ 
are respectively the electromagnetic gauge and  tensor fields\footnote {In this  work we adopt 
Gaussian natural units where $1\, {\rm GeV}^2 \simeq 5.13 \times 10^{19} \,$ G and  $e = 1/\sqrt{137}$.}. 
Since the model is non-renormalizable, we need to specify a regularization scheme. In this work we use a  
noncovariant cutoff regularization parametrized by $\Lambda$, within the magnetic field independent 
regularization scheme (MFIR). The MFIR scheme, originally formulated in terms of the 
proper-time regularization method~\cite{klimenko}, was recently reformulated~\cite {nosso1} 
using dimensional regularization by performing a sum over all Landau 
levels in the vacuum term. In this way, one is able to isolate the divergencies into a term 
that has the form of the zero magnetic field vacuum energy and thereby can be renormalized 
in standard fashion. The MFIR was recently employed in the problems of magnetized 
color superconducting cold matter~\cite{norberto,PRDdyana}, where its advantages, such as 
the avoidance of unphysical oscillations, are fully discussed. Other interesting application of MFIR scheme can be found in~\cite{mfirmpi0,mfirplb}, where the properties of magnetized neutral mesons were studied. Within this regularization scheme, 
the cutoff $\Lambda$, the coupling $G$ and the current quark mass $m$ represent free parameters 
which are fixed~\cite{klevansky,buballa} by fitting the vacuum values of the pion mass 
$m_{\pi}$, pion decay constant $f_{\pi}$ and quark condensate 
$\langle {\bar \psi}_f \psi_f \rangle$. 
 
In the MFA, the NJL pressure\footnote{Note that in this work we are concerned only with the longitudinal 
components of the pressure, sound velocity, etc. For simplicity they will be denoted as pressure, 
sound velocity, etc.} in the presence of a magnetic field can be expressed 
as a sum of quasi-particle and condensate contributions~\cite{nosso1,prdmarcus}: 
\begin{equation}
P = \frac{B^2}{2} + P_u + P_d - \frac{(M - m)^2}{4G} ,
\label{press}
\end{equation}
where $B^2/2$ comes from the first term in Eq.~(\ref{njl2}), and
each of the remaining terms can be written as a sum of three terms
($f=u,d$):
\begin{eqnarray}
P_f &=& P_f^{vac} + P_f^{mag} + P_f^{Tmag} ,  \\[0.25true cm]
\langle {\bar \psi}_f \psi_f \rangle  &=& \langle {\bar \psi}_f \psi_f \rangle^{vac}
+ \langle {\bar \psi}_f \psi_f \rangle^{mag}
+ \langle {\bar \psi}_f \psi_f \rangle^{Tmag}, 
\end{eqnarray}
with the quasi-particle terms given by 
\begin{eqnarray} 
\hspace{-0.5cm}
P^{vac}_{f} &=& \frac{N_c M^4}{8\pi^2} \left[ \frac{\epsilon_\Lambda \Lambda^3}{M^4}
\left(1 +  \frac{\epsilon^2_\Lambda}{\Lambda^2} \right )  
- \ln \left ( \frac{\Lambda + \epsilon_\Lambda}{M} \right )\right],
\label{Pvac} \\[0.25cm]
\hspace{-0.5cm}P^{mag}_f &=& \frac {N_c (|q_f| B)^2}{2 \pi^2} \biggl[ 
\frac{x^2_f}{4} - \frac{x_f}{2} (x_f-1) \ln x_f 
\nonumber \\
&& + \, \zeta^{\prime}(-1,x_f) \biggr],
\label{Pmag} \\[0.25cm]
\hspace{-0.5cm}P^{Tmag}_f &=& T \sum_{k=0}^{\infty} \alpha_k \frac {|q_f| B N_c }{2 \pi^2} \nonumber \\
&\times& \int_{-\infty}^{+\infty} dp \, \ln \left\{1 + \exp[-(E_f/T)] \right \} ,
\label{PmuB}
\end{eqnarray}

The quark condensates are given by
\begin{eqnarray}
\langle {\bar \psi}_f \psi_f \rangle^{vac} &=& -\frac{ M N_c }{2\pi^2} \left [
\Lambda \, \epsilon_\Lambda -
 {M^2} \ln \left ( \frac{\Lambda+ \epsilon_\Lambda}{{M }} \right ) \right ] ,
\label{Cvac}\\[0.25cm] 
\langle {\bar \psi}_f \psi_f \rangle^{mag}
&=& -\frac{ M |q_f| B N_c }{2\pi^2}\biggl[ \ln \Gamma(x_f)  -\frac {1}{2} \ln (2\pi) 
\nonumber \\
&& + \, x_f - \frac{1}{2} \left ( 2 x_f-1 \right )\ln (x_f) \biggr] ,
\label{Cmag} \\[0.25cm]
\langle {\bar \psi}_f \psi_f \rangle^{Tmag}&=&
\sum_{k=0}^{\infty} \alpha_k \frac{ M |q_f| B N_c }{2 \pi^2}
\int_{-\infty}^{+\infty} dp \; \frac{n(E_f)}{E_f} 
\label{MmuB} ,
\end{eqnarray}
where $\Gamma(x_f)$ is usual gamma function, and the other quantities appearing in these equations 
are given by 
\begin{eqnarray}
\epsilon_\Lambda &=& \left(\Lambda^2 + M^2 \right)^{1/2}, \\[0.25cm]
E_f &=& \left(p^2 + M^2 + 2 |q_f| B k\right)^{1/2}, \\[0.25cm] 
x_f &=& \frac{M^2}{2 |q_f| B}, \\[0.25cm]
n(E_f) &=& \frac{1}{1+\exp(E_f /T)} , \\[0.25cm]
\zeta^{\prime}(-1,x_f) &=& \frac{d\zeta(z,x_f)}{dz}\bigg|_{z=-1},
\end{eqnarray}
where $\zeta(z,x_f)$ is the Riemann-Hurwitz zeta function. To take further derivatives of
this function, as well as for numerical purposes, it is useful to use the following  
representation~\cite {BB}  
\begin{eqnarray}
\hspace{-0.5cm}\zeta^{\prime}(-1,x_f) &=& \zeta^{\prime}(-1,0) \nonumber \\
&& + \, \frac{x_f}{2}\left[ x_f-1-\ln(2\pi) 
+ \psi^{(-2)}(x_f) \right],
\label{zeta'}
\end{eqnarray}
where $\psi^{(m)}(x_f)$ is  the $m$-th polygamma function and the $x_f$ independent constant 
is $\zeta^{\prime}(-1,0)=-1/12$. In the sum in Eq.~(\ref{MmuB}), $k$ represents the Landau levels. 
In addition, $M$ represents the MFA effective quark mass, which is solution of  the gap equation:  
\begin{equation}
M = m - 2G \sum_{f=u}^{d}\langle {\bar \psi}_f \psi_f \rangle .
\label{gapeqnjl}
\end{equation}

Notice that although the quark condensate for the flavors $u$ and $d$ in the presence 
of a magnetic field  are different due to their different electric charges, the masses 
of the $u$ and $d$ constituent quarks are equal to each other since we 
work here in isospin-symmetric limit, $m_u=m_d=m${\textemdash}for details,  
see Ref.~\cite {prdmarcus}. Finally note that the  term $B^2/2$ in Eq. (\ref {press})
does not contribute to the normalized pressure $P_N(T,B) = P(T,B) - P(0,B)$ 
(see Ref.~\cite {nosso1} for further details).

At vanishing densities, the energy density $\epsilon$ is defined as $\epsilon = - P_N + Ts $, 
where  $s$ is the entropy density,  $s=\partial P_N / \partial T$. Other thermodynamical 
observables such as  the interaction 
measure, $\Delta$,  the specific heat, $c_v$, the velocity of sound, $c_s^2$,  and the 
magnetization, $\cal M$, which contain valuable  information on the role played by  
the magnetic field on the onset of chiral transition, will also be investigated here. 
They are defined as follows
\begin{equation}
c_v = \left(\frac{\partial \varepsilon}{\partial T}\right)_v, \hspace{0.25cm}
\Delta = \frac{\varepsilon - 3P_N}{T^4}\;, \hspace{0.25cm}
c_s^2 = \left(\frac{\partial P_N}{\partial \varepsilon}\right)_v ,
\end{equation}
and 
\begin{equation}
{\cal M} = \frac{dP_N}{dB} .
\end{equation}

%%%%%%%%%%%%%%%%%%%%%%%%%%%%%%%%%%%%%
\section{Thermo-magnetic NJL Coupling }

\begin{figure*}[htb]
\begin{center}
\includegraphics[width=0.5\linewidth,angle=0]{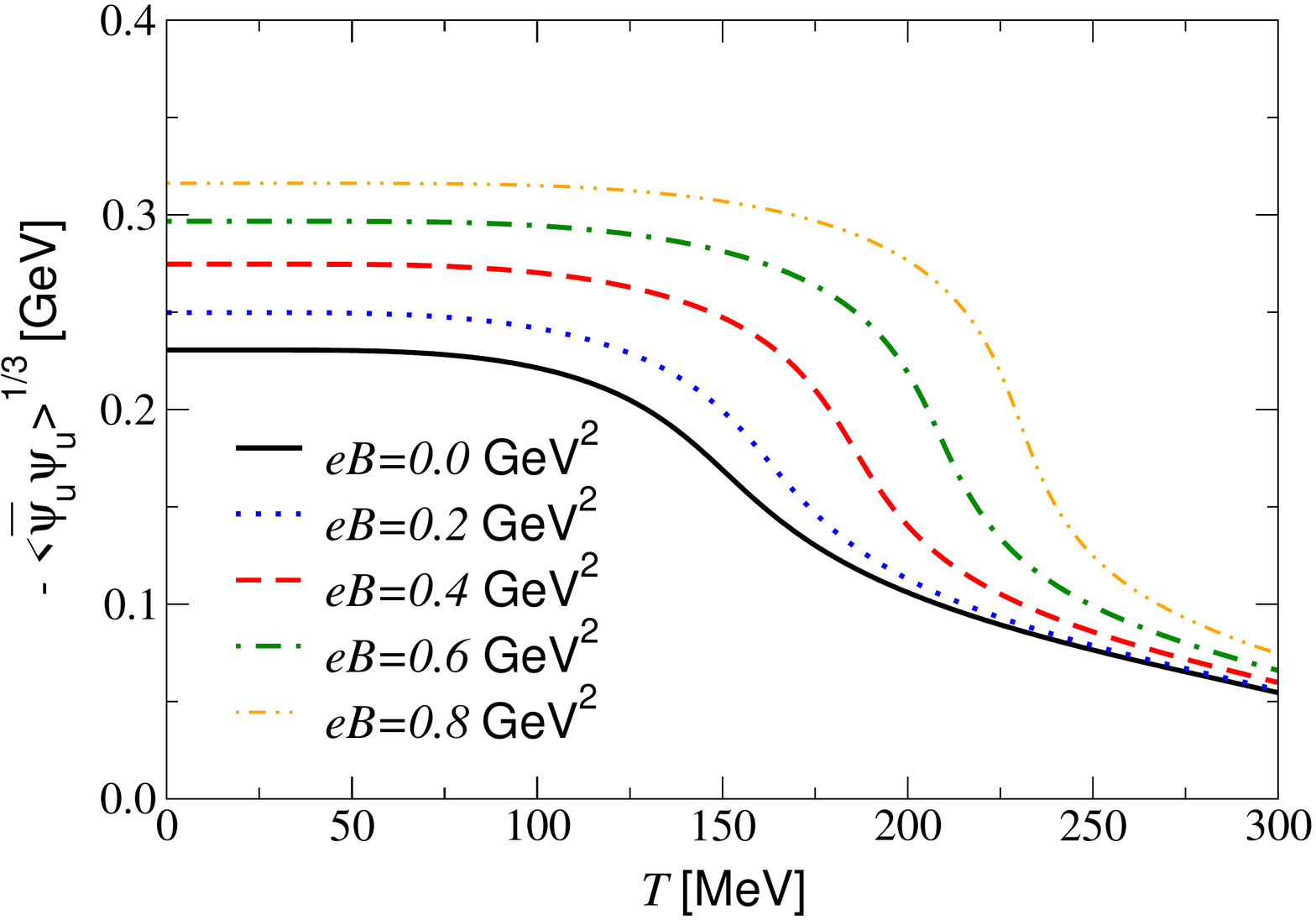}\hspace*{0.0cm}
\includegraphics[width=0.5\linewidth,angle=0]{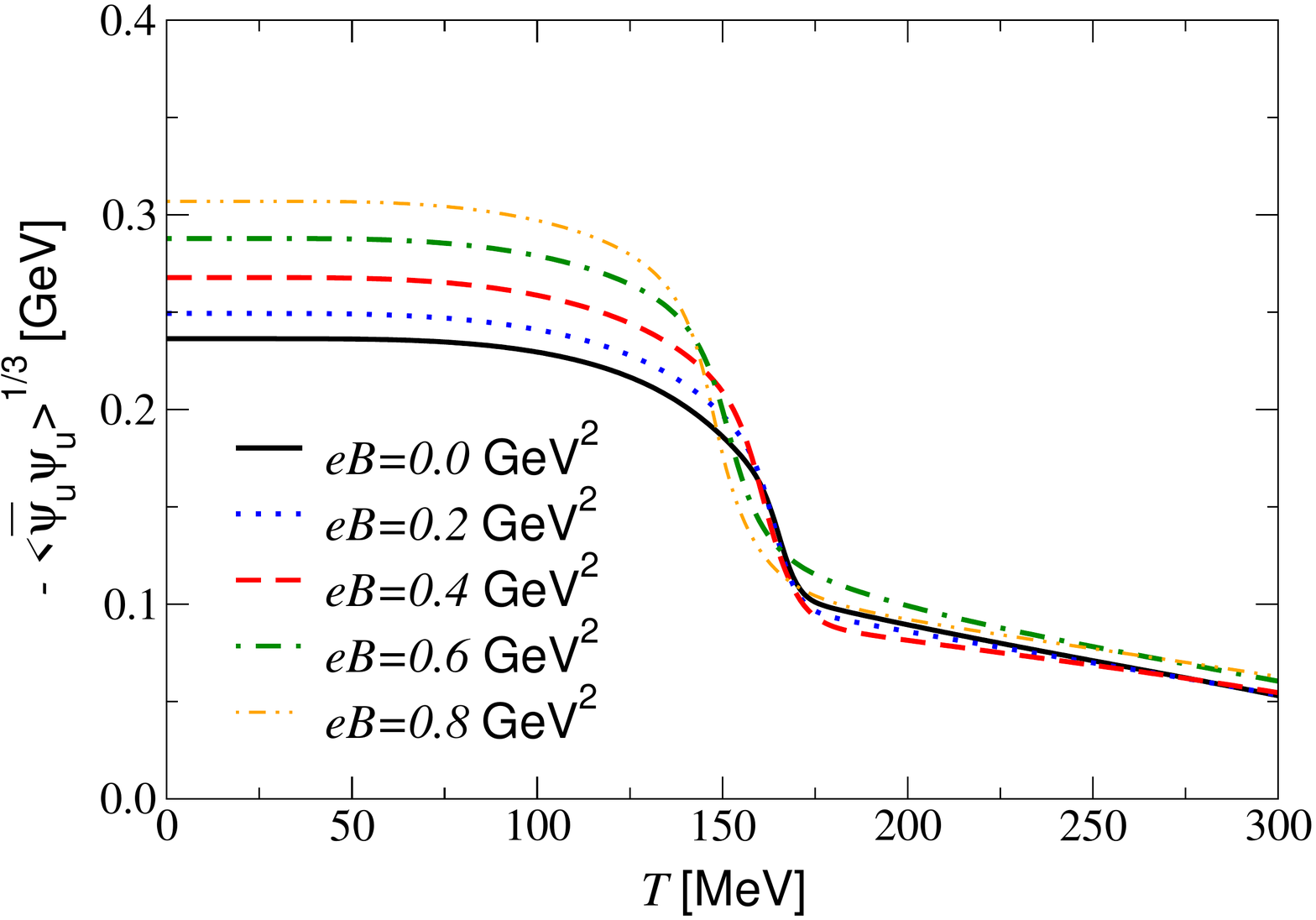}
\vspace*{0.0cm}
\includegraphics[width=0.5\linewidth,angle=0]{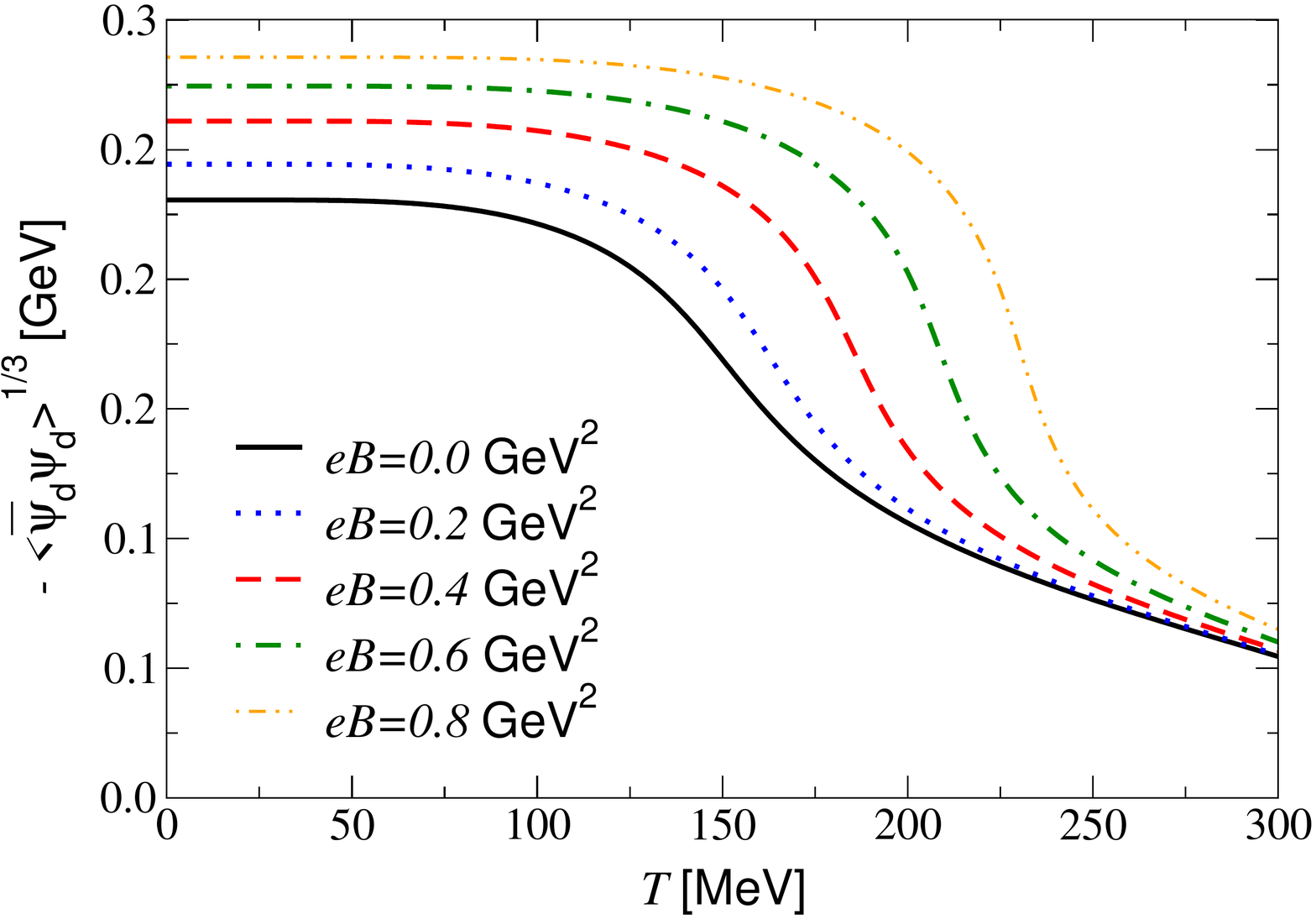}\hspace*{0.0cm}
\includegraphics[width=0.5\linewidth,angle=0]{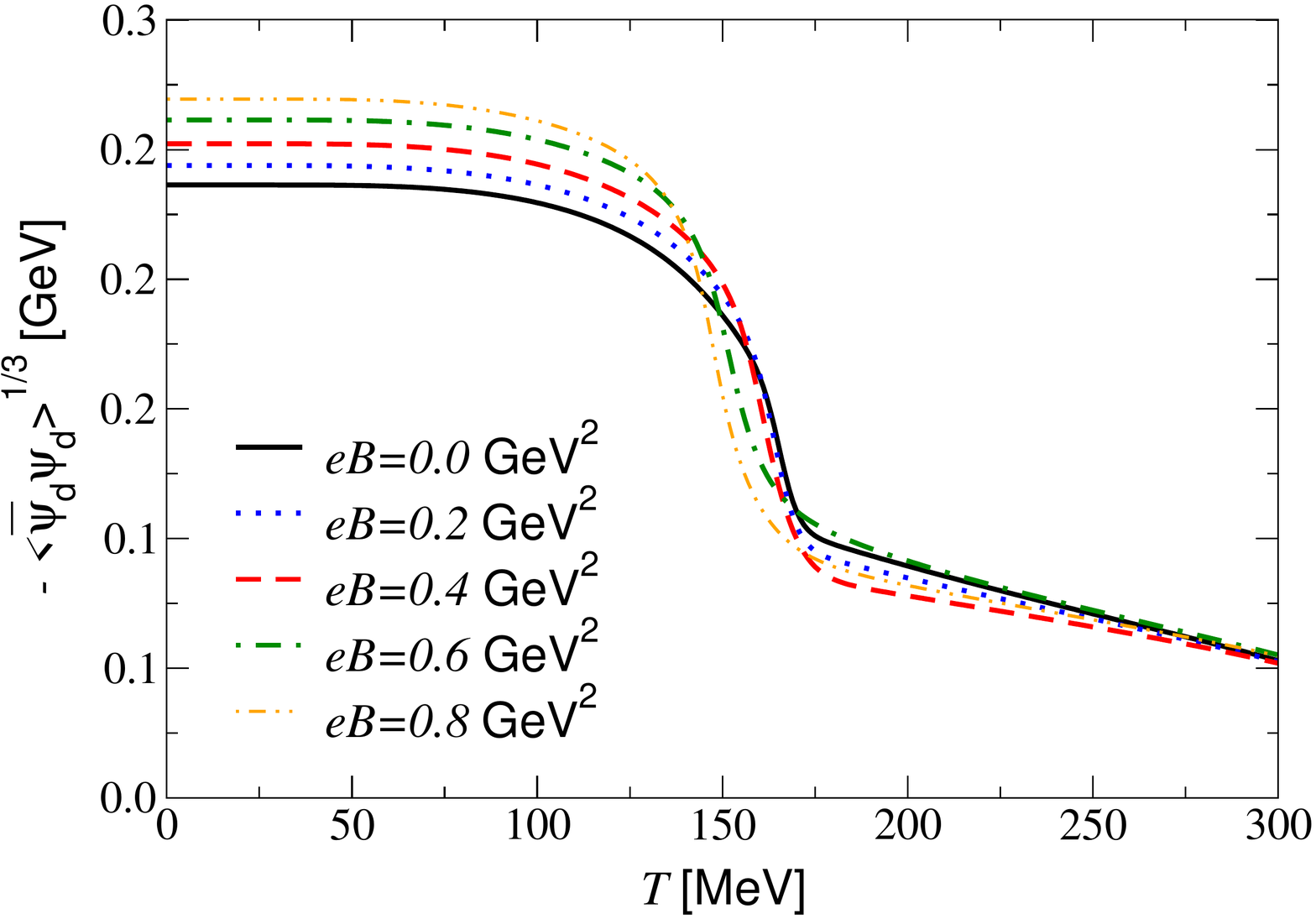}
\caption{Quark condensate for flavors $u$ and $d$ as functions 
of temperature for different values of the magnetic field for $G$ (left) and 
$G(B,T)$ (right).}
\label{fig1}
\end{center}
\end{figure*}

We start describing  the fitting procedure used to obtain the thermo-magnetic 
dependence of the NJL coupling constant. Our strategy is to reproduce with the model
the lattice results of Ref.~\cite{fodor} for the quark condensate average, 
$ (\Sigma_u + \Sigma_d)/2$. In the lattice calculation, the condensates are 
normalized in a way which is reminiscent of Gell-Mann--Oakes--Renner relation 
(GOR), $2m \langle \bar \psi_f\psi_f \rangle = m_\pi^2 f_\pi^2 + \dots$, as
\begin{equation}
\Sigma_{f}(B,T) = \frac{2m}{m_\pi^2 f_\pi^2}\left[\langle \bar \psi_f\psi_f \rangle_{BT} 
- \langle \bar \psi_f\psi_f \rangle_{00}\right]+1 ,
\label{sigmalatt}
 \end{equation}
with $ \langle \bar \psi_f\psi_f \rangle_{00}$ representing the quark condensates at $T=0$ 
and $B=0$. In order to fit the lattice results, the other physical quantities appearing in  
Eq.~(\ref {sigmalatt}) should be those of Ref.~\cite {fodor}; namely, $m_\pi =135\; {\rm MeV}$, 
$f_\pi= 86\; {\rm MeV}$, and $m=5.5 \;{\rm MeV}$ so that, by invoking the GOR relation,  
one can use the LQCD value $\langle \bar \psi_f\psi_f \rangle_{00}^{1/3}=-230.55 \, {\rm MeV}$.
Therefore, as far as Eq.~(\ref{sigmalatt}) is concerned, only 
$\langle \bar \psi_f\psi_f \rangle_{BT}$ is to be evaluated with the NJL model. As we show below, 
the NJL predictions for the in vacuum scalar condensate are numerically very close to those
obtained with the LQCD simulations so that the above value for $\langle\bar\psi_f\psi_f \rangle_{00}$
can be safely used in Eq.~(\ref {sigmalatt}) without introducing important uncertainties.

\begin{figure*}[htb]
\begin{center} 
\includegraphics[width=0.5\linewidth,angle=0]{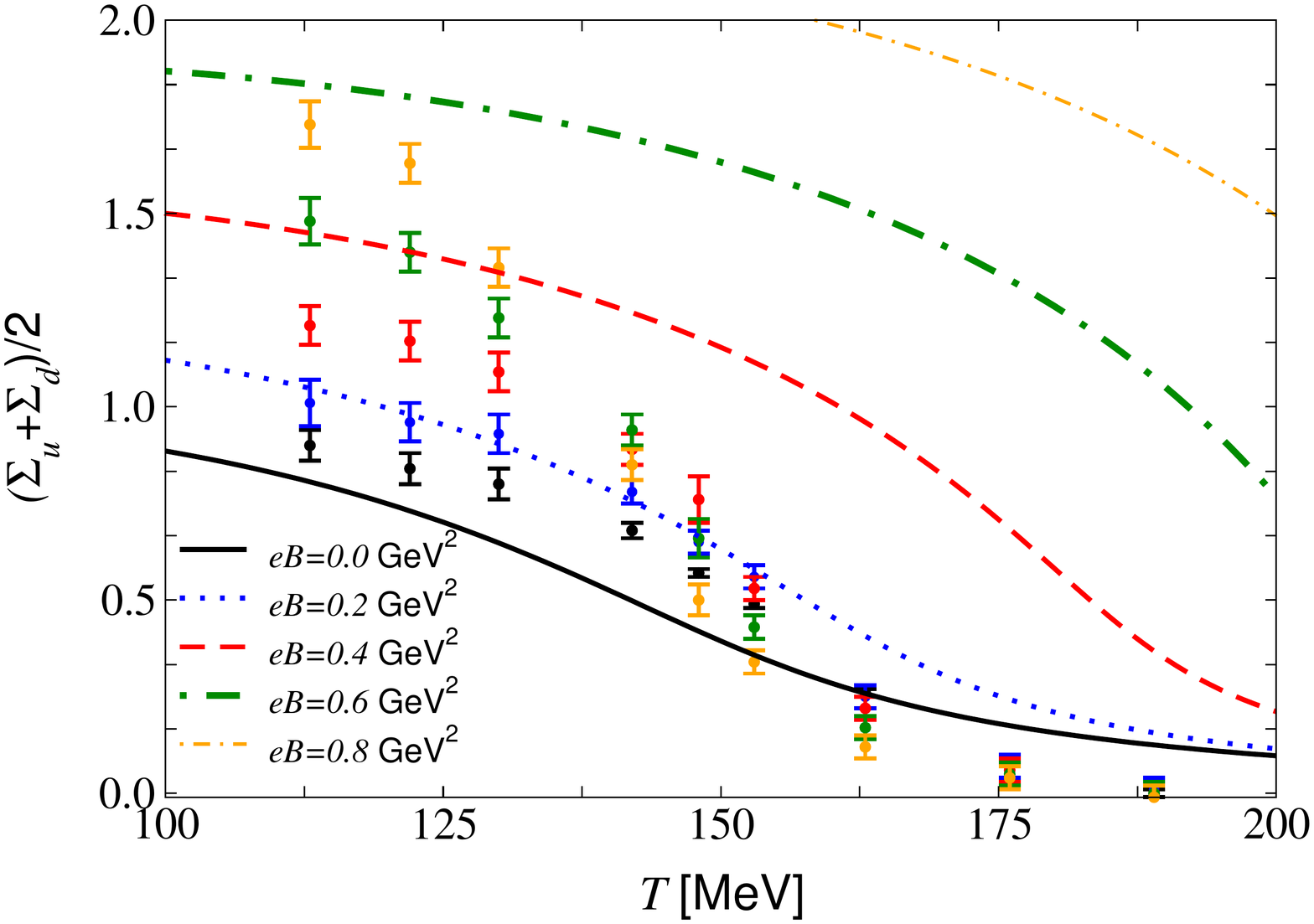}\hspace*{0.0cm} 
\includegraphics[width=0.5\linewidth,angle=0]{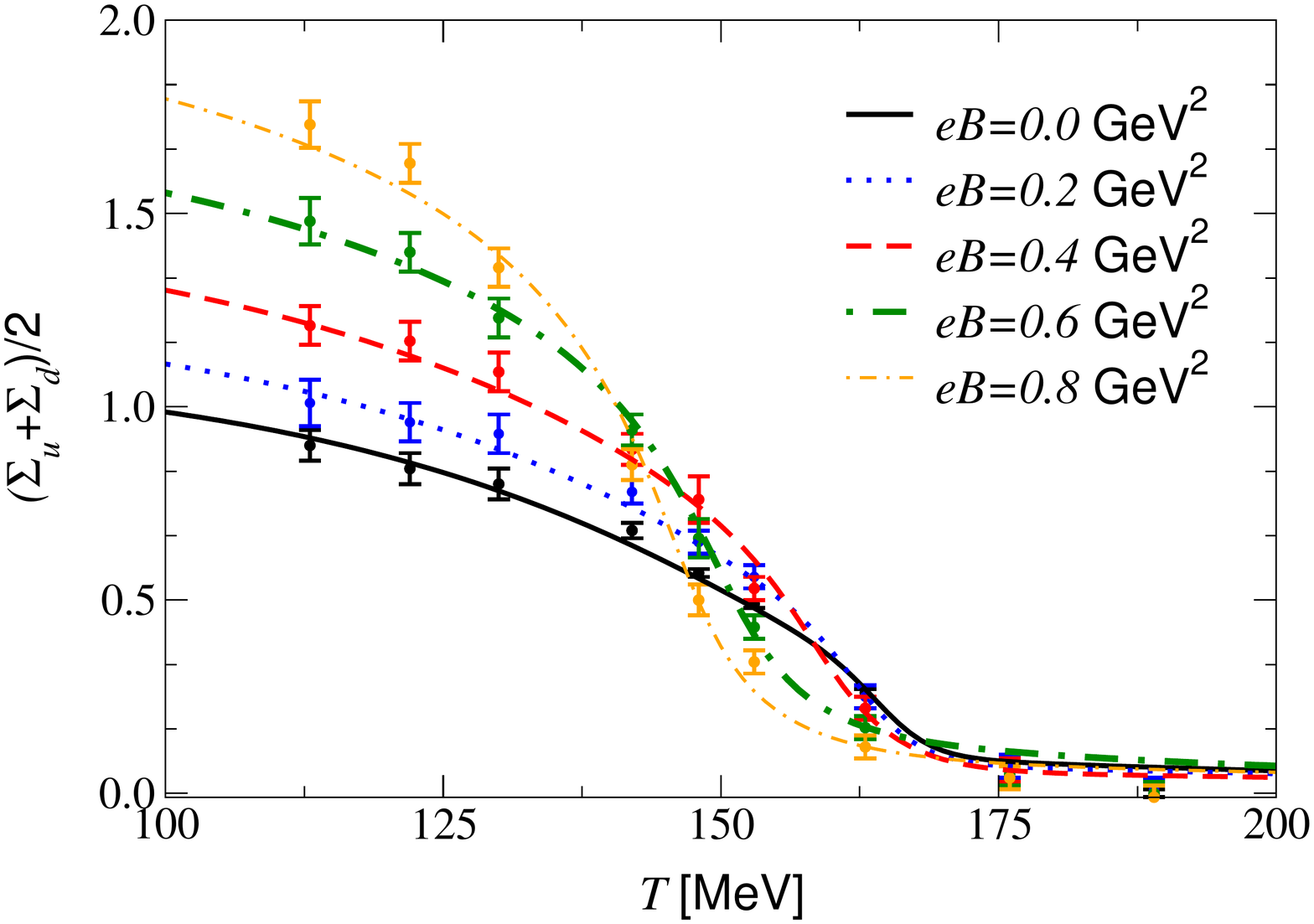}
\vspace*{0.0cm}
\includegraphics[width=0.5\linewidth,angle=0]{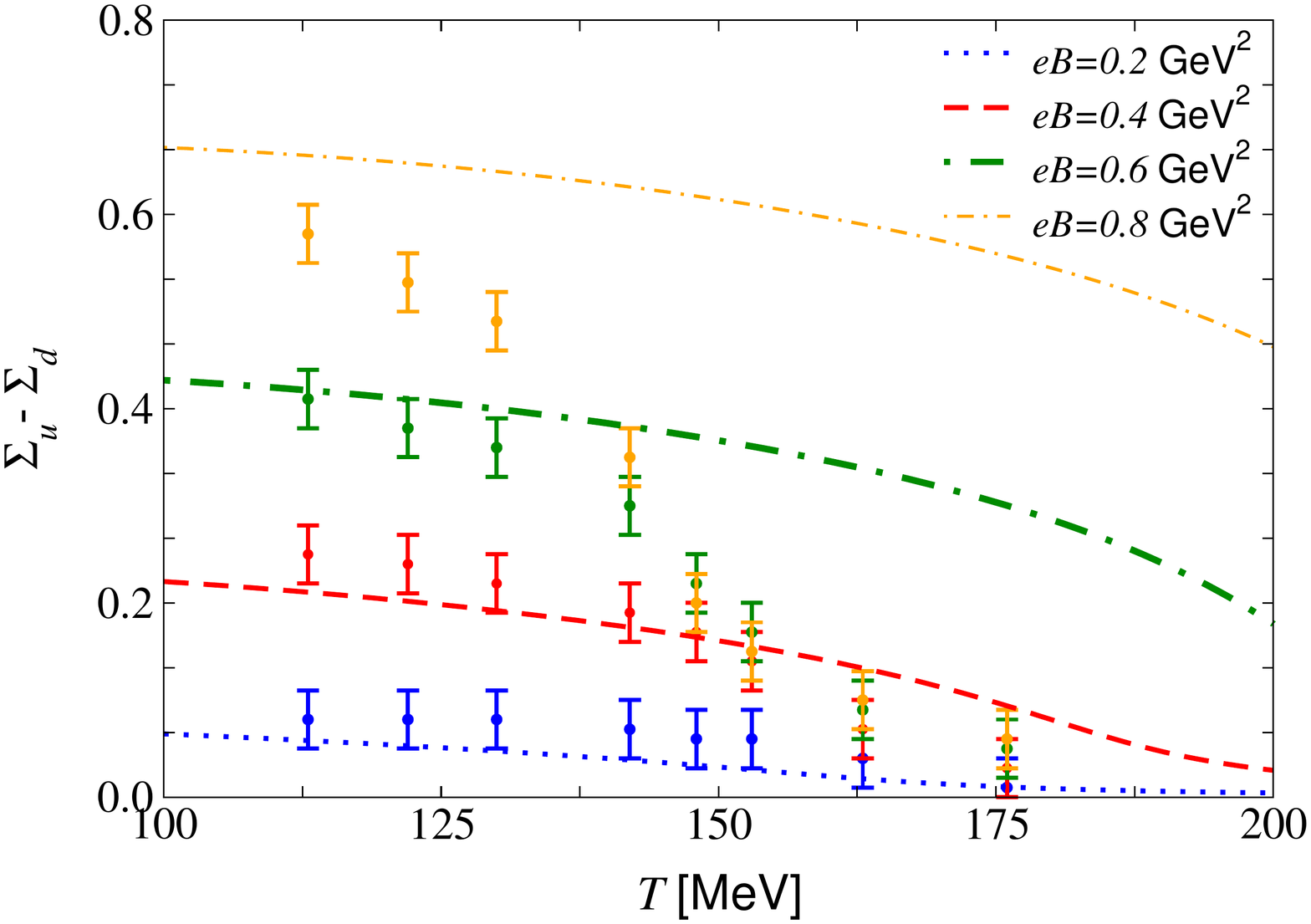}\hspace*{0.0cm} 
\includegraphics[width=0.5\linewidth,angle=0]{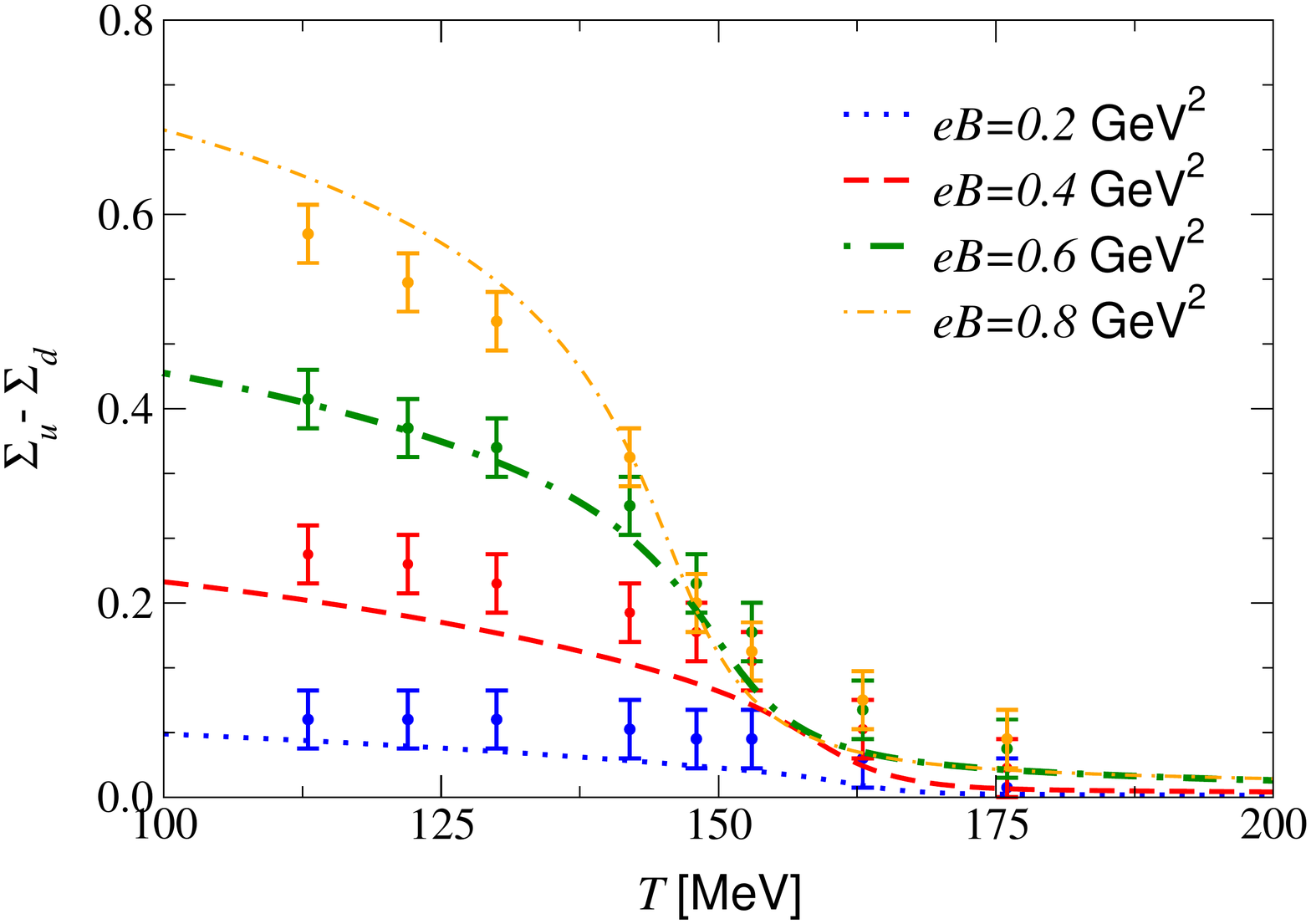}
\caption{Condensate average and difference as functions 
of temperature for different values of the magnetic field for $G$ (left) and 
$G(B,T)$ (right). Data from Ref.~\cite{fodor}.}
\label{fig2}
\end{center}
\end{figure*}

The LQCD results of Ref.~\cite{fodor} were obtained at $T=0$ and at high~$T$, with no data 
points between $T=0$ and $T=113~{\rm MeV}$. Therefore, recalling that the discrepancies 
between lattice results and effective models appear in the region where chiral symmetry is 
partially restored (crossover), it seems therefore reasonable to fit the NJL coupling 
constant within this region and then to extrapolate the results to zero temperature if needed.

A good finite temperature fit to the lattice data for the average $(\Sigma_u + \Sigma_d)/2$ 
can be obtained by using the following interpolation formula for NJL coupling constant:
\begin{eqnarray}
G(B,T) = c(B)\left[1-\frac{1}{1+ e^{\beta(B) \left[T_a(B) - T\right]}}\right]+s(B) .
\label{ourGBT}
\end{eqnarray}
Note that the parameters $c,s,\beta$ and $T_a$ depend only on the magnetic field; their values
are shown in Table~\ref{glatnjl}. Remark also that Eq.~(\ref {ourGBT}) does not necessarily 
require the knowledge of $G(0,0)$, but one still needs $\Lambda$ and $m$ which in this work are 
taken at standard values, $\Lambda = 0.650~{\rm GeV}$ and $m = 5.5~{\rm MeV}$. This particular
expression is taken simply for convenience; any other form that fits the data is expected
to give the same qualitative results for thermodynamical quantities in the appropriate
$B$ and $T$ range. 

\begin{table}
\caption{Values of the fitting parameters in Eq.~(\ref{ourGBT}). Units are in appropriate 
powers of GeV.}
\label{glatnjl}
\begin{tabular}{ccccc}
\hline\noalign{\smallskip}
$eB$ & $c$ &  $T_a$ & $s$   & $\beta$  \\ \hline\noalign{\smallskip}
0.0    &    0.900 &  0.168   &  3.731   &  40.000 \\\noalign{\smallskip}
0.2    &    1.226  &  0.168   &  3.262   &  34.117 \\\noalign{\smallskip}
0.4    &    1.769  &  0.169   &  2.294   &  22.988 \\\noalign{\smallskip}
0.6    &    0.741  &  0.156   &  2.864   &  14.401 \\\noalign{\smallskip}
0.8    &    1.289  &  0.158   &  1.804   &  11.506 \\\noalign{\smallskip}
\hline
\end{tabular}
\end{table}

Figures~\ref{fig1} and ~\ref{fig2} display the results for combinations of the quark condensates: the $u$ 
and $d$ condensates, their sum and difference. In the left panels of the figures, the condensates 
are evaluated with a $T-$ and $B-$independent coupling $G$ that fits the lattice results for 
the average $(\Sigma_u + \Sigma_d)/2$ in vacuum, $G = 4.50373~{\rm GeV}^{-2}$; in the right 
panels, the condensates are calculated with the coupling $G(B,T)$ of Eq.~(\ref{ourGBT}), 
with the fitting parameters given in Table~\ref{glatnjl}.

The figures clearly show that 
the NJL model is able to capture the sharp decrease around the crossover temperature 
of the lattice results for the average {\em and} difference of the condensates only when 
the coupling $G(B,T)$ is used; when using the $T-$ and $B-$independent coupling~$G$, a rather smooth 
behavior for these quantities is obtained. We~have not attempted to obtain a 
$G(B,T)$ that gives a best fit for both $(\Sigma_u + \Sigma_d)/2$  and $\Sigma_u - \Sigma_d$, 
but one sees that the model nevertheless gives a very reasonable description of the latter. 
Although here we are not particularly concerned with the results at $T=0$, for the sake of
completeness we mention that an extrapolation of the fit to $T=B=0$ gives
$G(0,0) = 4.6311~{\rm GeV}^{-2}$. Such a coupling leads to 
$\langle {\bar \psi}_f \psi_f\rangle^{1/3}_{00}=-236.374~{\rm MeV}$, which differs 
only by a few percent from the value calculated with $G$. This small discrepancy is due to the
fact that we have attempted to obtain a good fit with a limited number of parameters 
of the lattice data at high temperatures only, where more data are available. 

%%%%%%%%%%%%%%%%%%%%%%%%%%%%%%%%%%%%
\section{Thermodynamical quantities} 
\label{results}

\begin{figure*}[htb]
\begin{center}
\includegraphics[width=0.5\linewidth,angle=0]{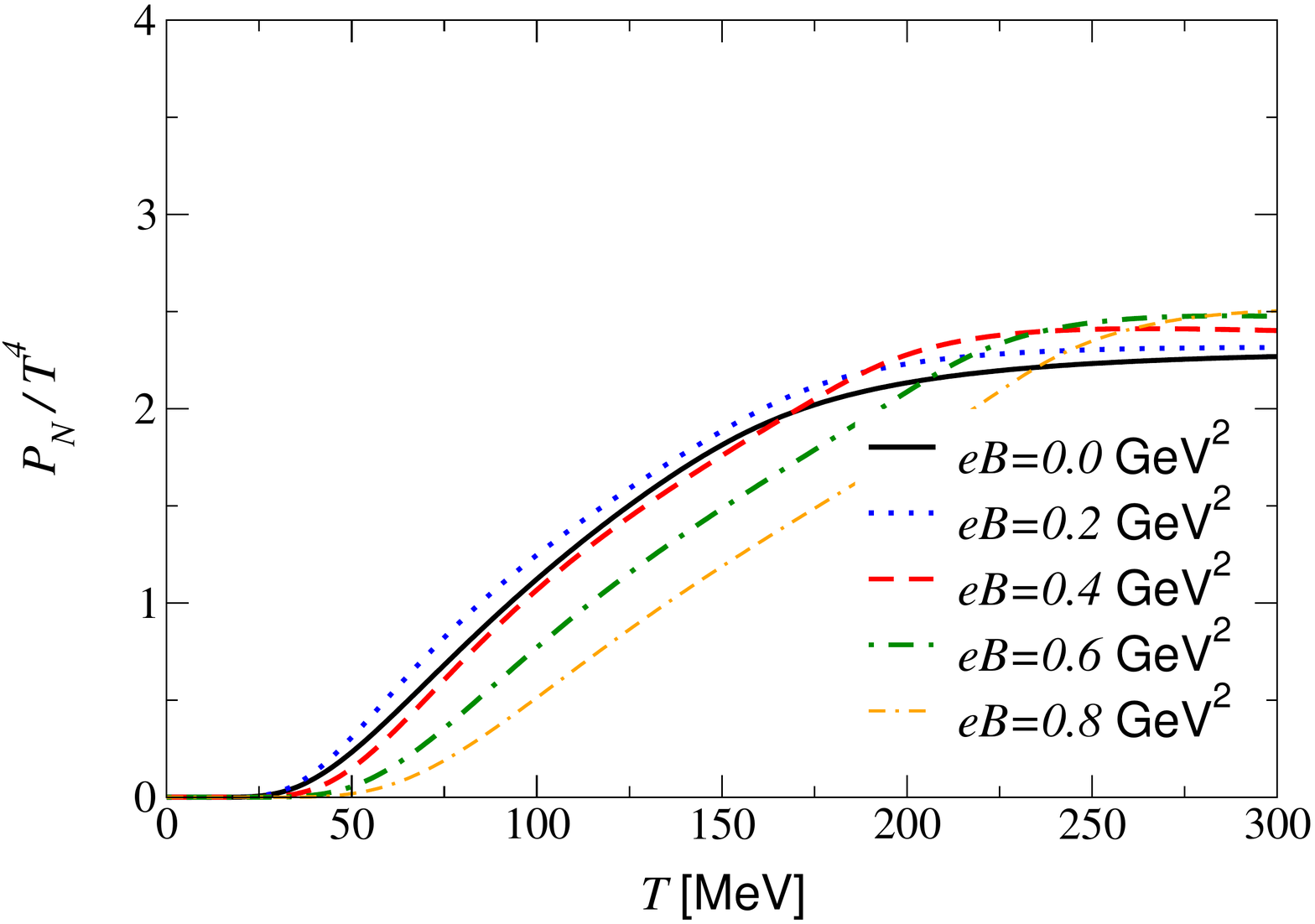}\hspace*{0.0cm}
\includegraphics[width=0.5\linewidth,angle=0]{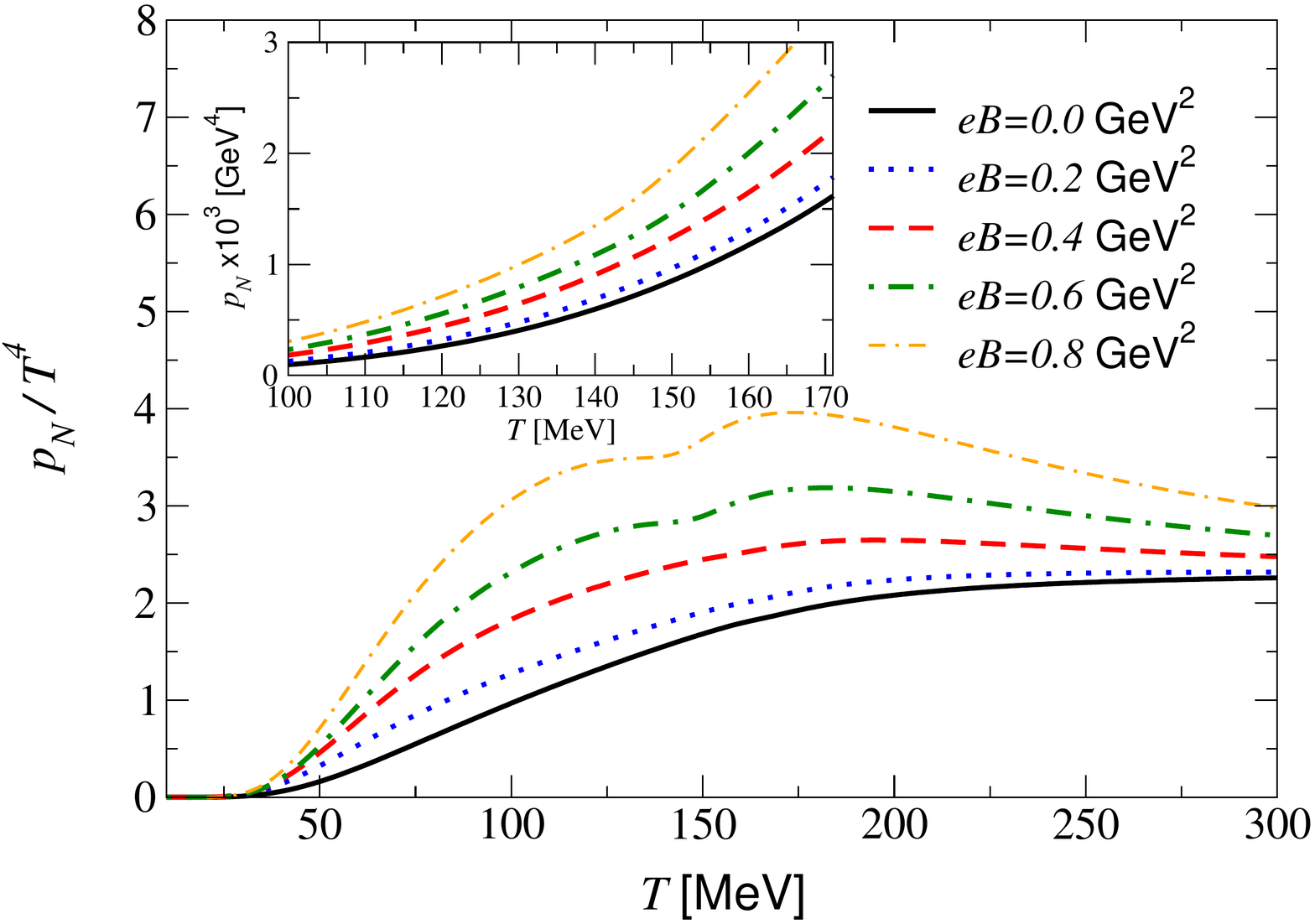}
\vspace*{0.0cm}
\includegraphics[width=0.5\linewidth,angle=0]{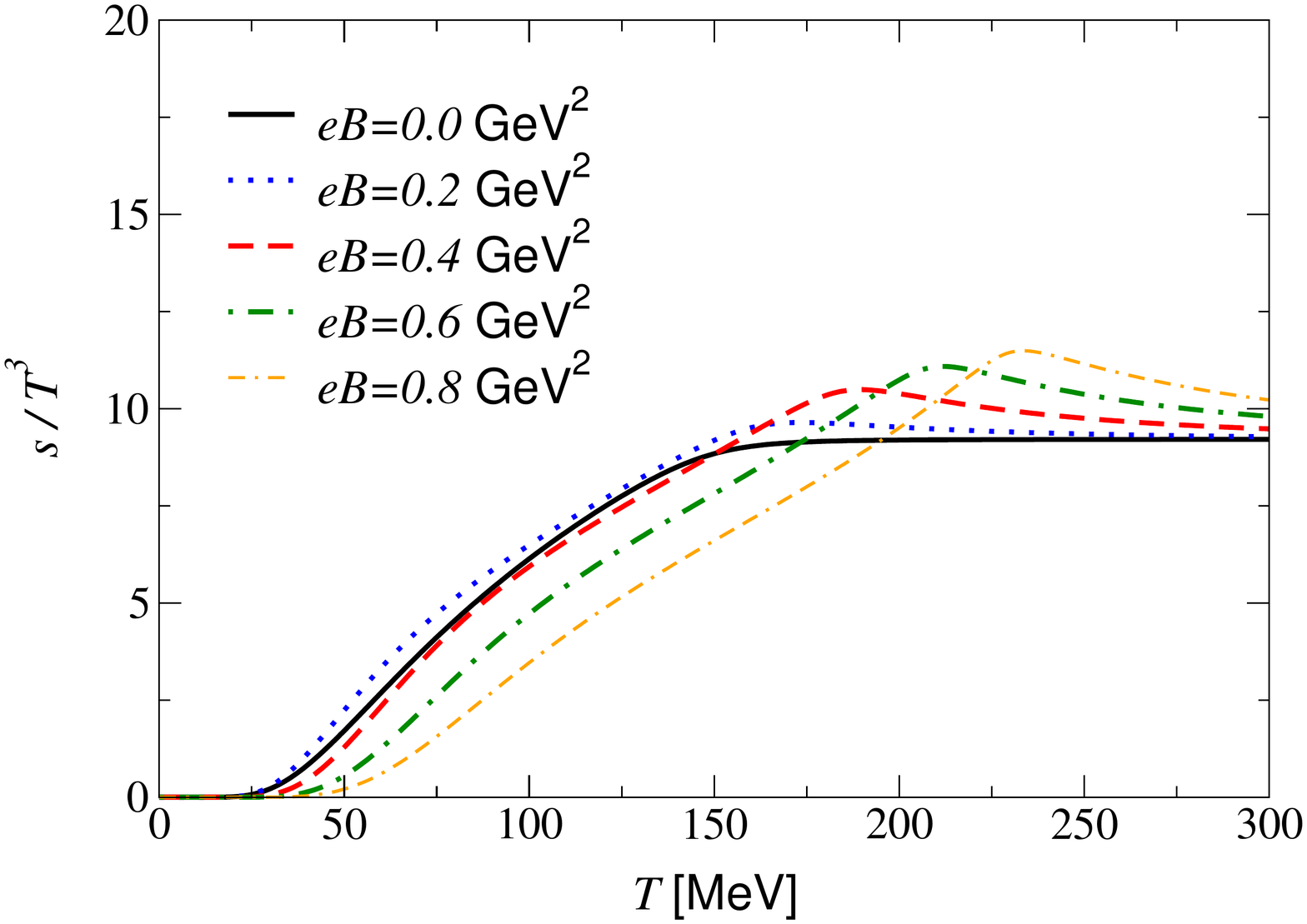}\hspace*{0.0cm}
\includegraphics[width=0.5\linewidth,angle=0]{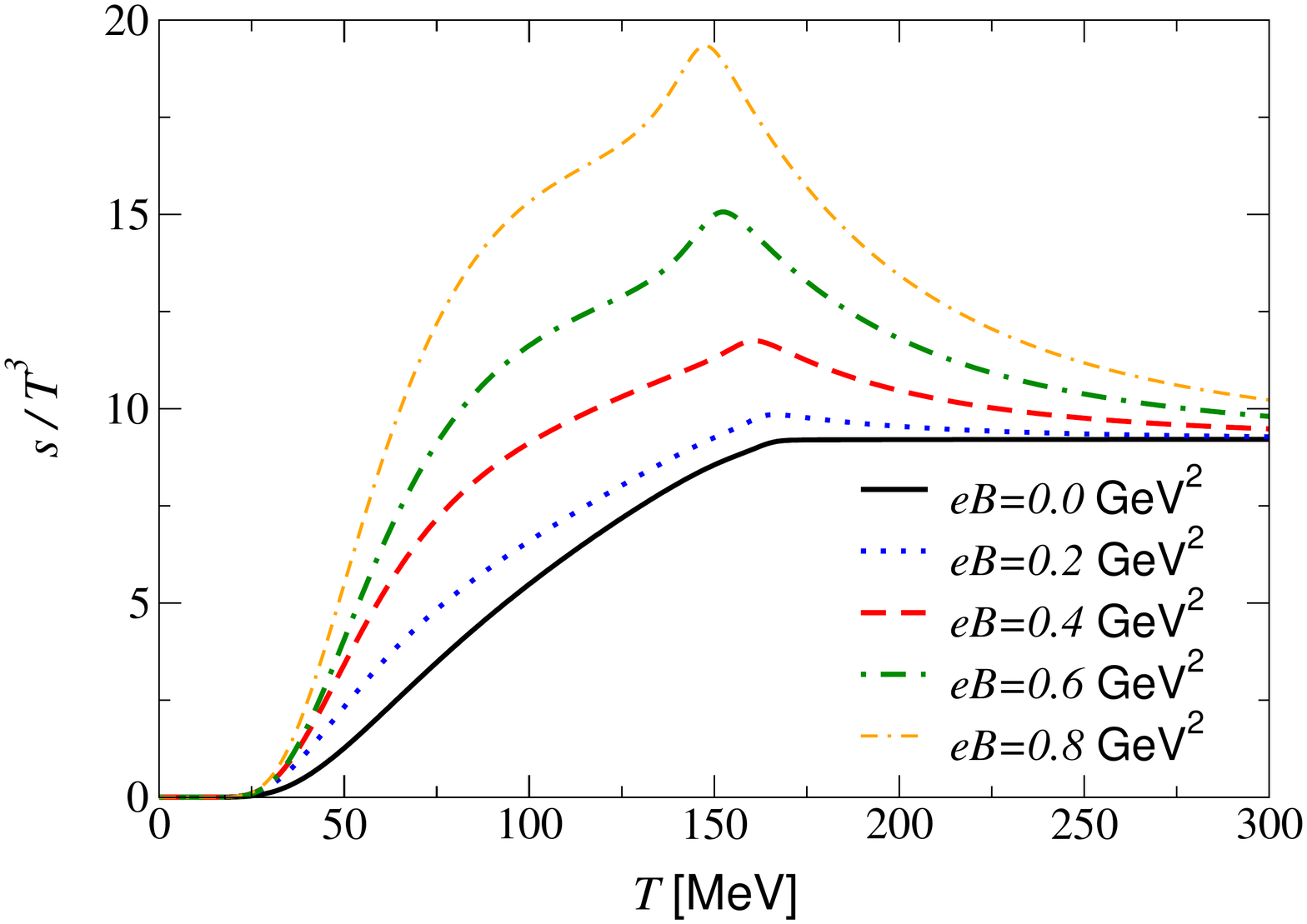}
\caption{Normalized pressure and entropy density as functions of temperature 
for different values of the magnetic field calculated with $G$ (left) and $G(B,T)$ (right).}
\label{fig3}
\end{center}
\end{figure*}

\begin{figure*}[htb]
\begin{center}
\includegraphics[width=0.5\linewidth,angle=0]{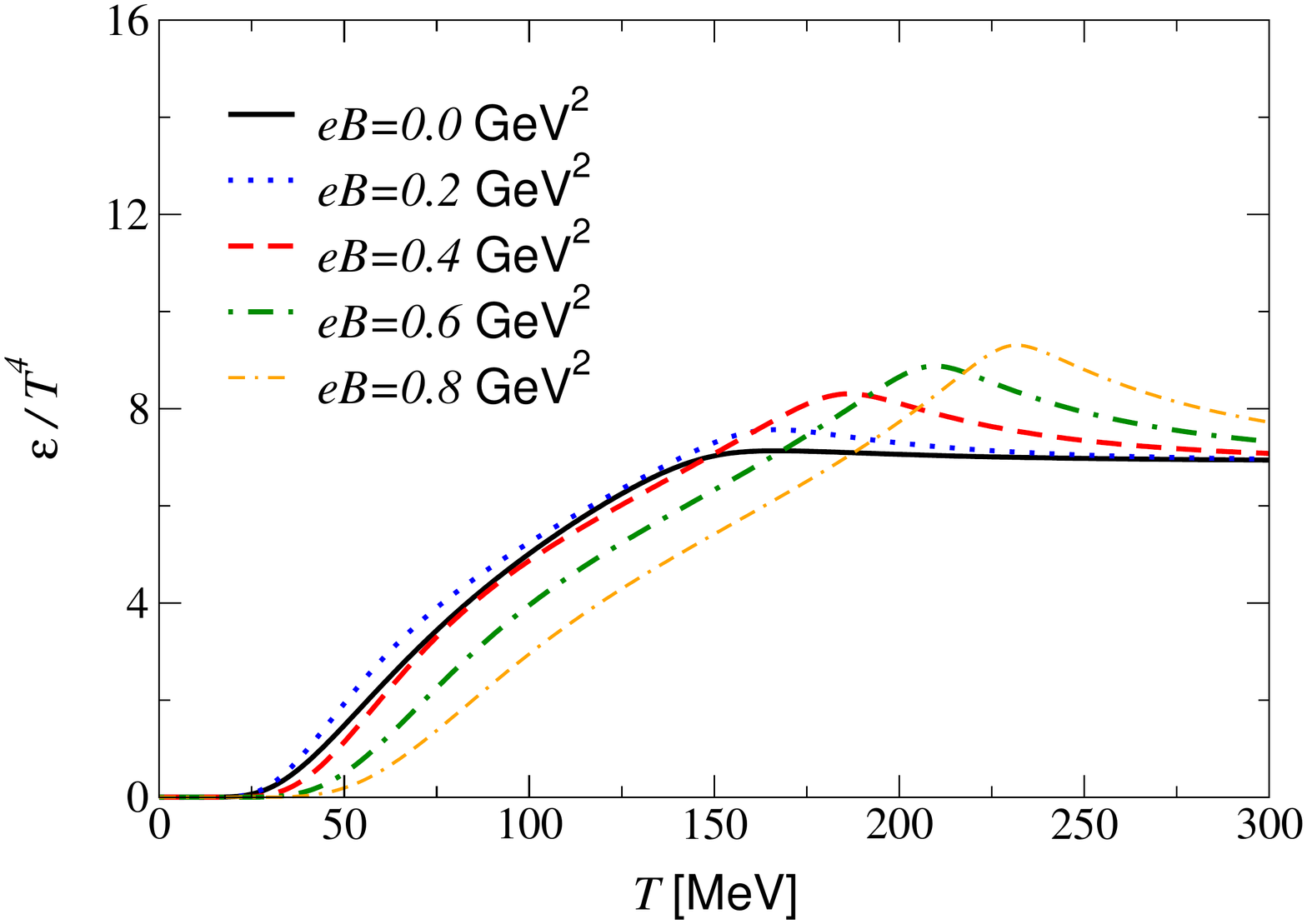}\hspace*{0.0cm}
\includegraphics[width=0.5\linewidth,angle=0]{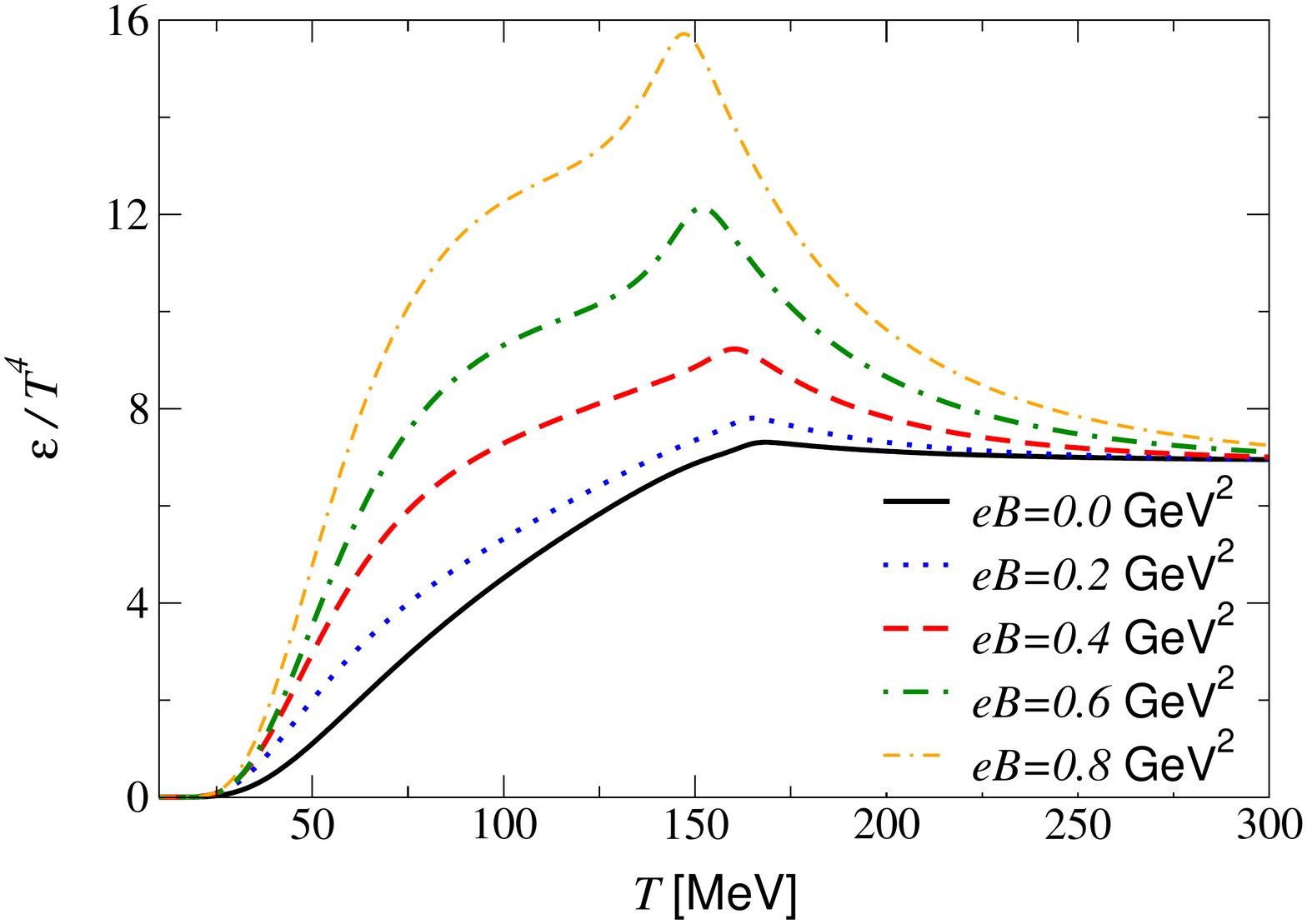}
\vspace*{0.0cm}
\includegraphics[width=0.5\linewidth,angle=0]{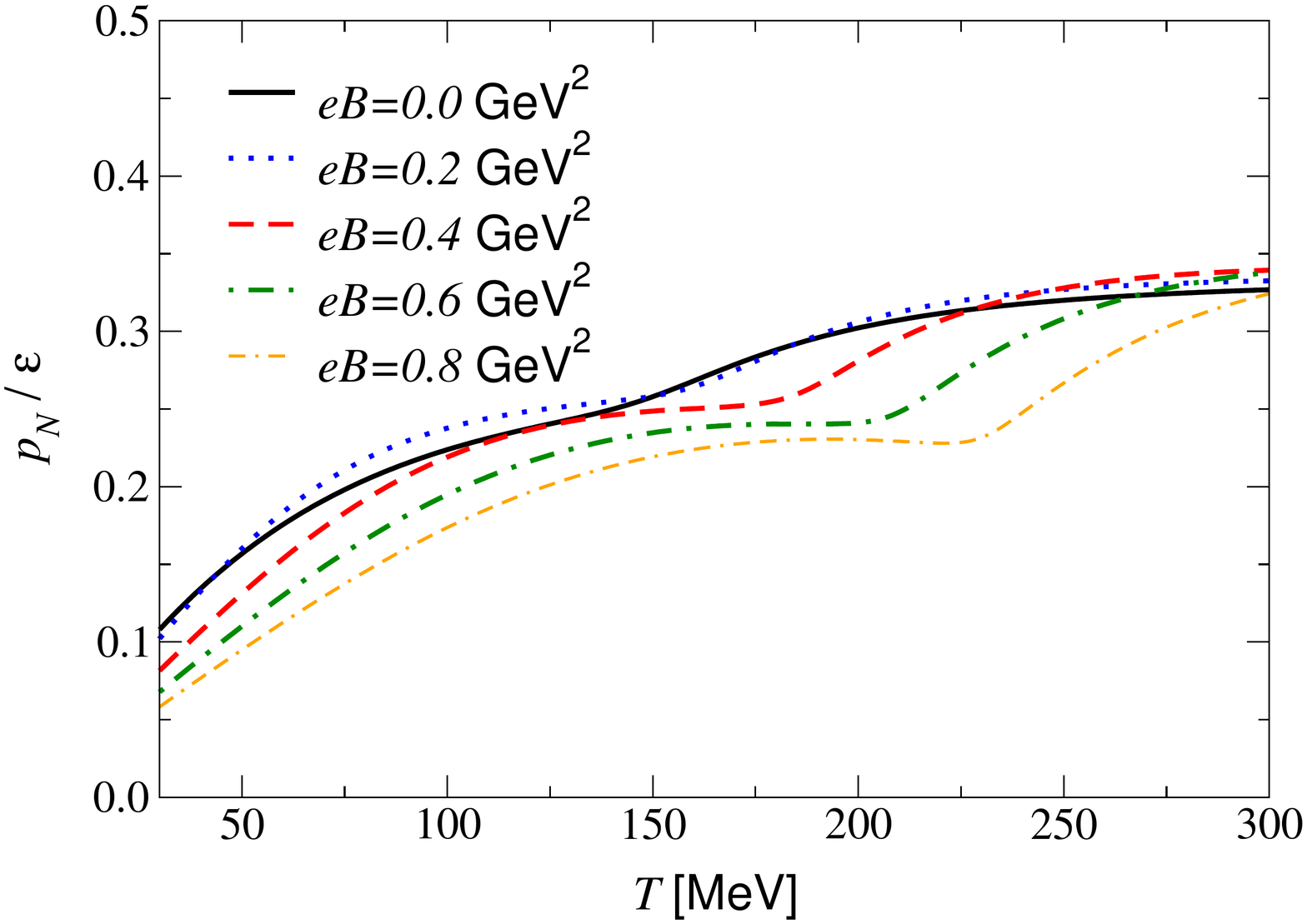}\hspace*{0.0cm}
\includegraphics[width=0.5\linewidth,angle=0]{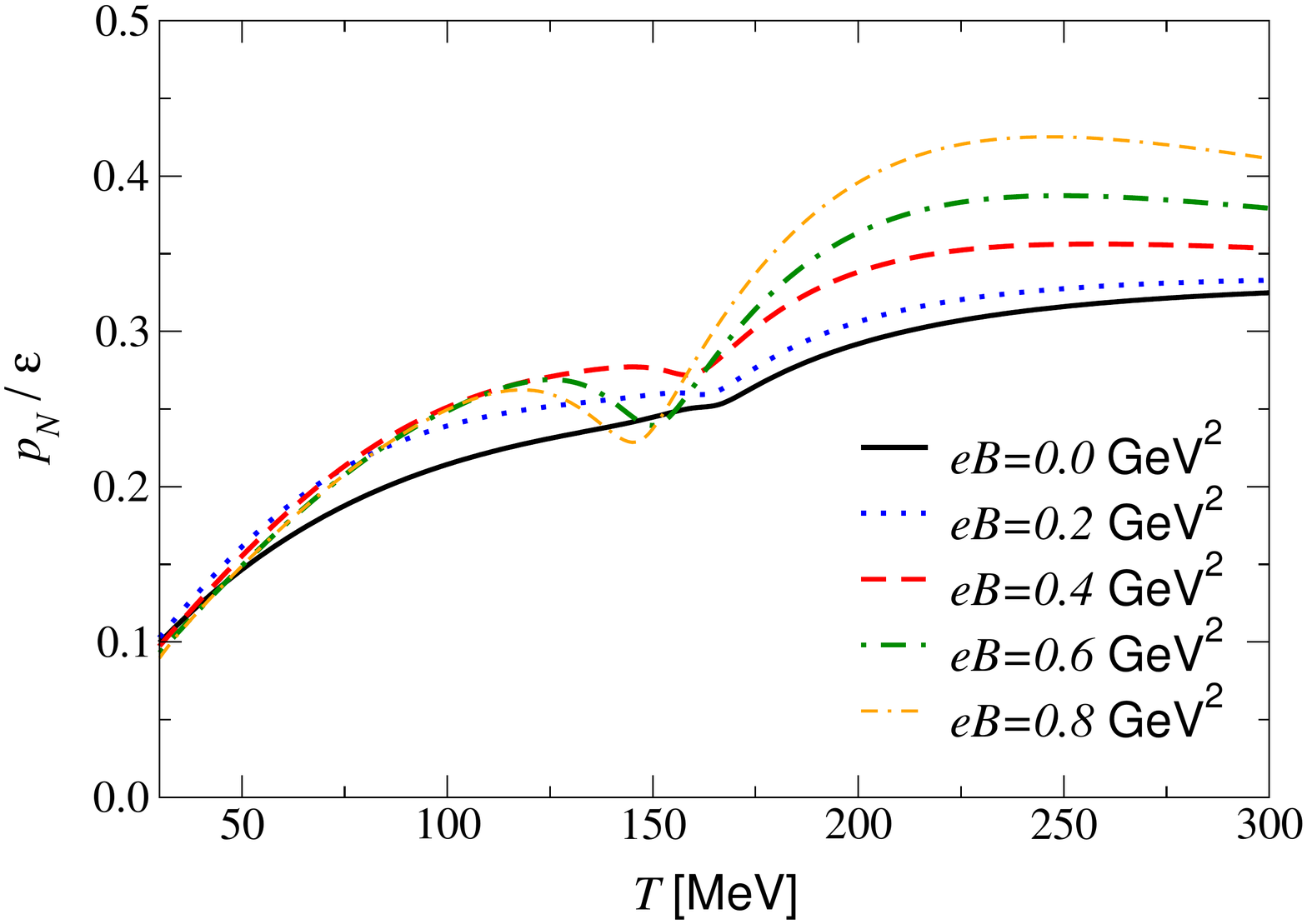}
\caption{Energy density and equation of state as functions of temperature 
for different values of the magnetic field calculated with $G$ (left) and $G(B,T)$ (right).}
\label{fig4}
\end{center}
\end{figure*}

In the present section we examine the predictions of the NJL model for the thermodynamical quantities 
of magnetized quark matter when the fitted coupling $G(B,T)$ is used. We start by considering the quantities that 
characterize the equation of state (EoS), such as the normalized pressure $P_N = P(T,B) - P(0,B)$, the 
entropy density $s$, the energy density $\cal E$, and the EoS parameter $P_N/{\cal E}$. 
These quantities are displayed in Figures~\ref {fig3} and~\ref {fig4} for the $T-$ and $B-$independent 
$G$ and $G(B,T)$ couplings. We note that the choice to present results for $P_N$, instead for 
$P(T,B)$ and $P(0,B)$ in separate, is for presentation convenience; $P_N$ vanishes at $T=0$ and the 
explicit dependence of the term $B^2/2$ cancels in the difference. 

A first observation is that $G(B,T)$ always predicts larger values for $P_N$, $s$, and $\cal E$ 
for a given temperature as the field strength increases as compared to the corresponding values 
obtained with a $T-$ and $B-$independent coupling $G$. Moreover, qualitative different $T$ dependences at high and low temperature values can be observed by comparing the two top panels of Fig. 3. Starting with the left panel, which corresponds to  $G$, one observes that the pressure value decreases as  $B$ increases at very low temperatures ($T  \lesssim 50 \, {\rm MeV}$) while an opposite trend is observed at high temperatures ($T  \gtrsim 250 \, {\rm MeV}$). One also observes that at intermediate temperatures ($T \simeq 200\, {\rm MeV}$) the pressure values predicted by increasingly higher fields oscillate around the $B=0$ value. On the other hand, the top right panel corresponding to $G(B,T)$ predicts that a  higher field  always leads to a higher pressure. The $G(B,T)$ coupling also predicts 
a more dramatic increase of the pressure around the chiral crossover; at $eB=0.8\; {\rm GeV}^2$, 
the pressure predicted with $G(B,T)$ is about twice the value at $B=0$, while the departure 
from the $B=0$ curve is more modest in the case when $G$ is considered.

The NJL predictions with a $G(B,T)$ for thermodynamical observables can be contrasted with recent 
lattice results~\cite{bali}. For example, the systematic increase of $P_N$ with $B$ is clearly 
observed in Fig.~5 of Ref.~\cite{bali} while the behavior of $P_N / \epsilon$ seen in Fig. 7 
of the same reference is similar to the one found in Fig.~\ref {fig3} and Fig.~\ref {fig4}
of the present paper. We remark that a different normalization has been used in Ref.~\cite {bali} 
and, at first sight, our results for $P_N/T^4$ seem to contradict those in the latter 
reference, but this is not so; the inset in Fig.~\ref{fig3} shows that, for a given $T$, 
the pressure always increases with~$B$, like in the left panel of Fig.~7 in Ref.~\cite {bali}. 
Our results for $s/T^3$ and ${\cal E}/T^4$ also predict a sharper transition and so the peaks 
are more pronounced than the ones which appear at fixed $G$. Even more important is the 
fact that, for increasing $B$, the peaks occur at lower temperatures, in a clear indication 
of IMC. Finally, notice that although the lattice calculations of Ref.~\cite{bali} are for 
$2+1$ flavors, a qualitative agreement can clearly be noticed.  We emphasize once more that 
the results with fixed $G$ present a clear discrepancy with the ones obtained within the  
LQCD simulations of Ref. \cite{bali}.

Before investigating other thermodynamical quant\-ities, let us recall that the
crossover temperature, or the pseudo-critical temperature $T_{pc}$, for which chiral symmetry 
is partially restored, is usually defined as the temperature at which the thermal 
susceptibility
\begin{eqnarray}
\chi_T &=& - m_{\pi}\frac{\partial \sigma}{\partial T}\,,  \\[0.25true cm]
\sigma &=& \frac{\langle\bar{\psi}_u\psi_u\rangle(B,T)+\langle\bar{\psi}_d\psi_d\rangle(B,T)}
{\langle\bar{\psi}_u\psi_u\rangle(B,0)
+\langle\bar{\psi}_d\psi_d\rangle(B,0)} ,
\label{chiT}
\end{eqnarray}
reaches a maximum. Note that we have followed the usual LQCD definition which includes 
the pion mass in the definition of $\chi_T$ to make it a dimensionless quantity. 
As in the previous section, we again consider $m_\pi=135\;~{\rm MeV}$. 

\begin{figure*}[htb]
\begin{center}
\includegraphics[width=0.5\linewidth,angle=0]{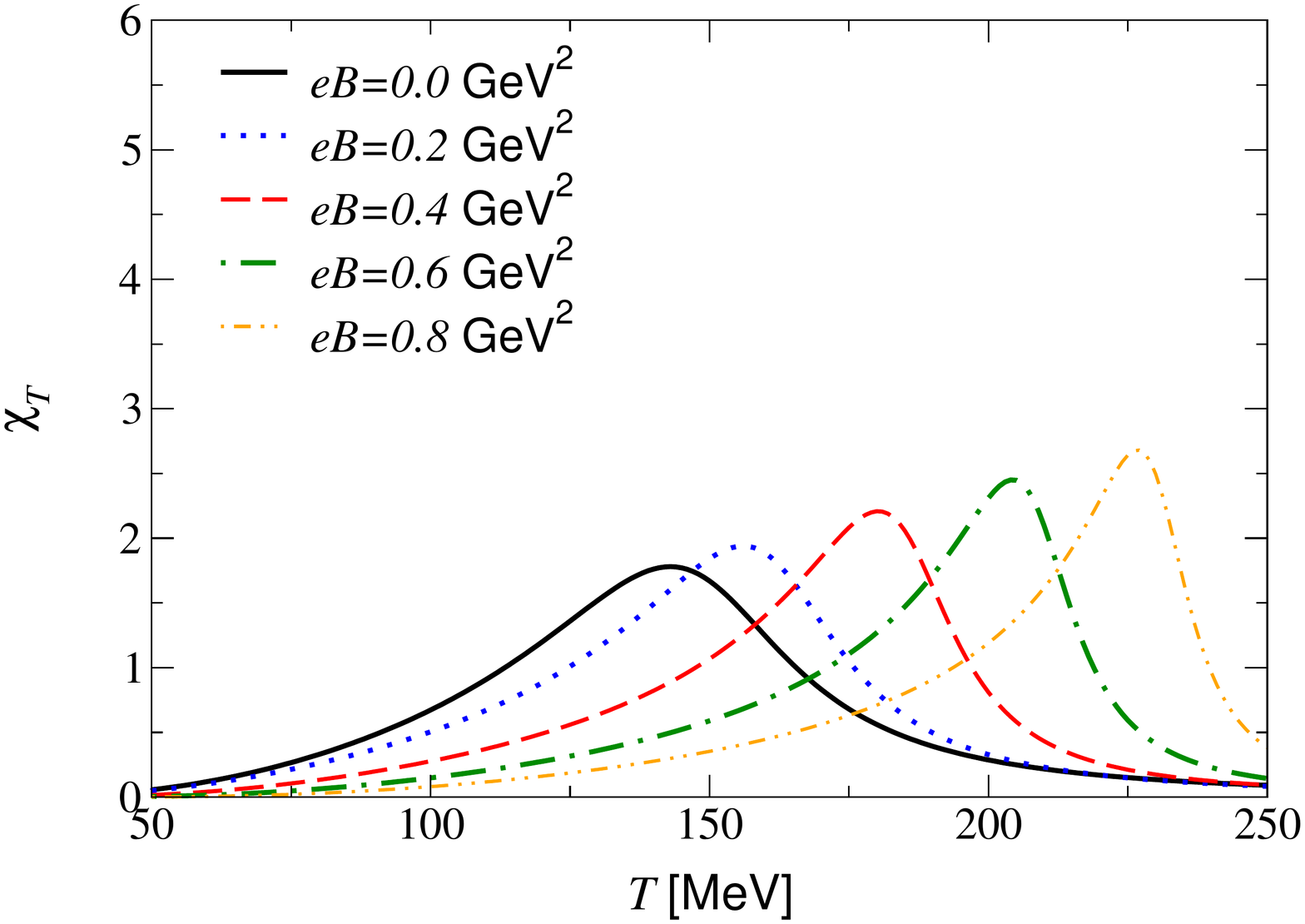}\hspace*{0.0cm}
\includegraphics[width=0.5\linewidth,angle=0]{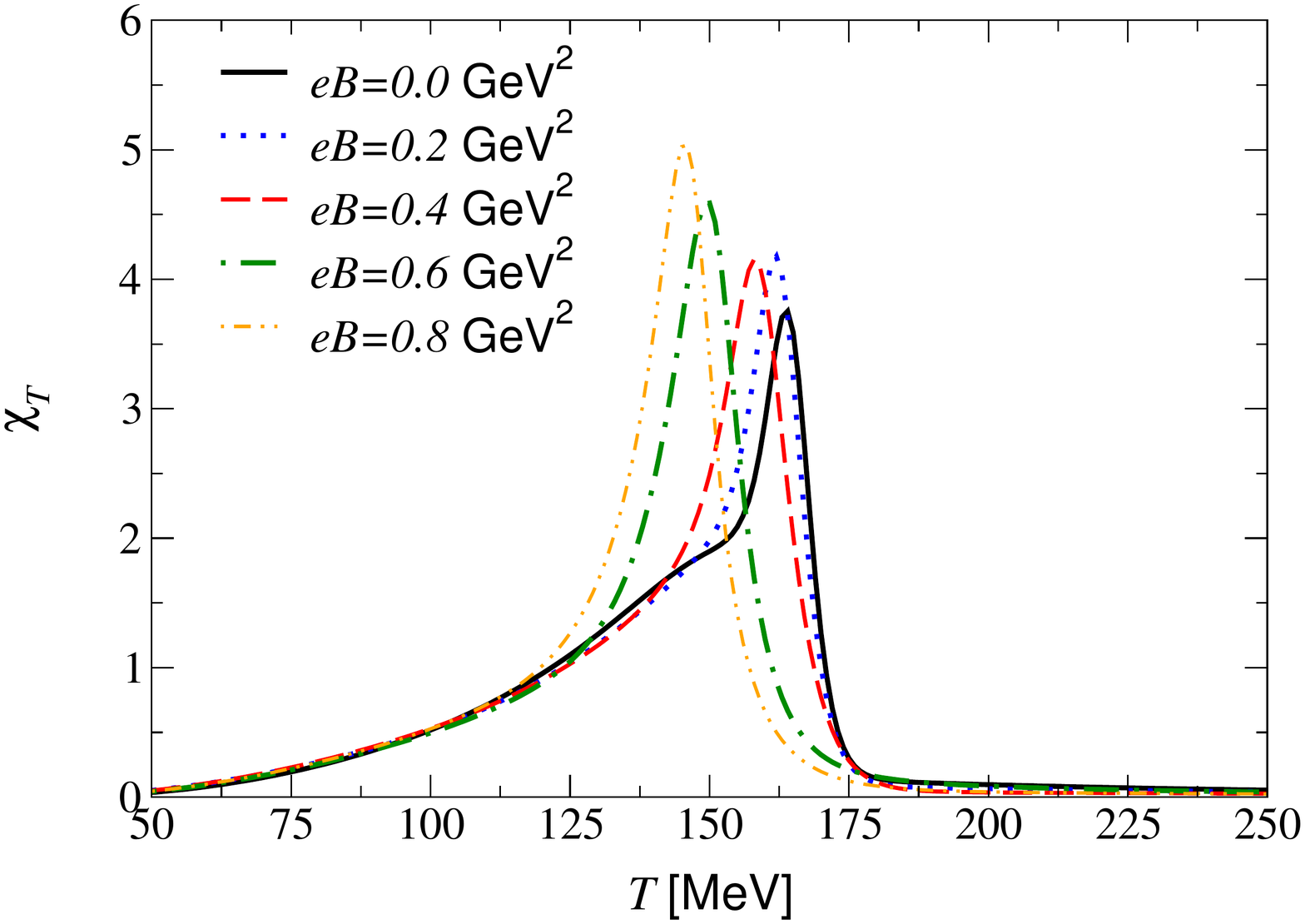}
\vspace*{0.00cm}
\includegraphics[width=0.5\linewidth,angle=0]{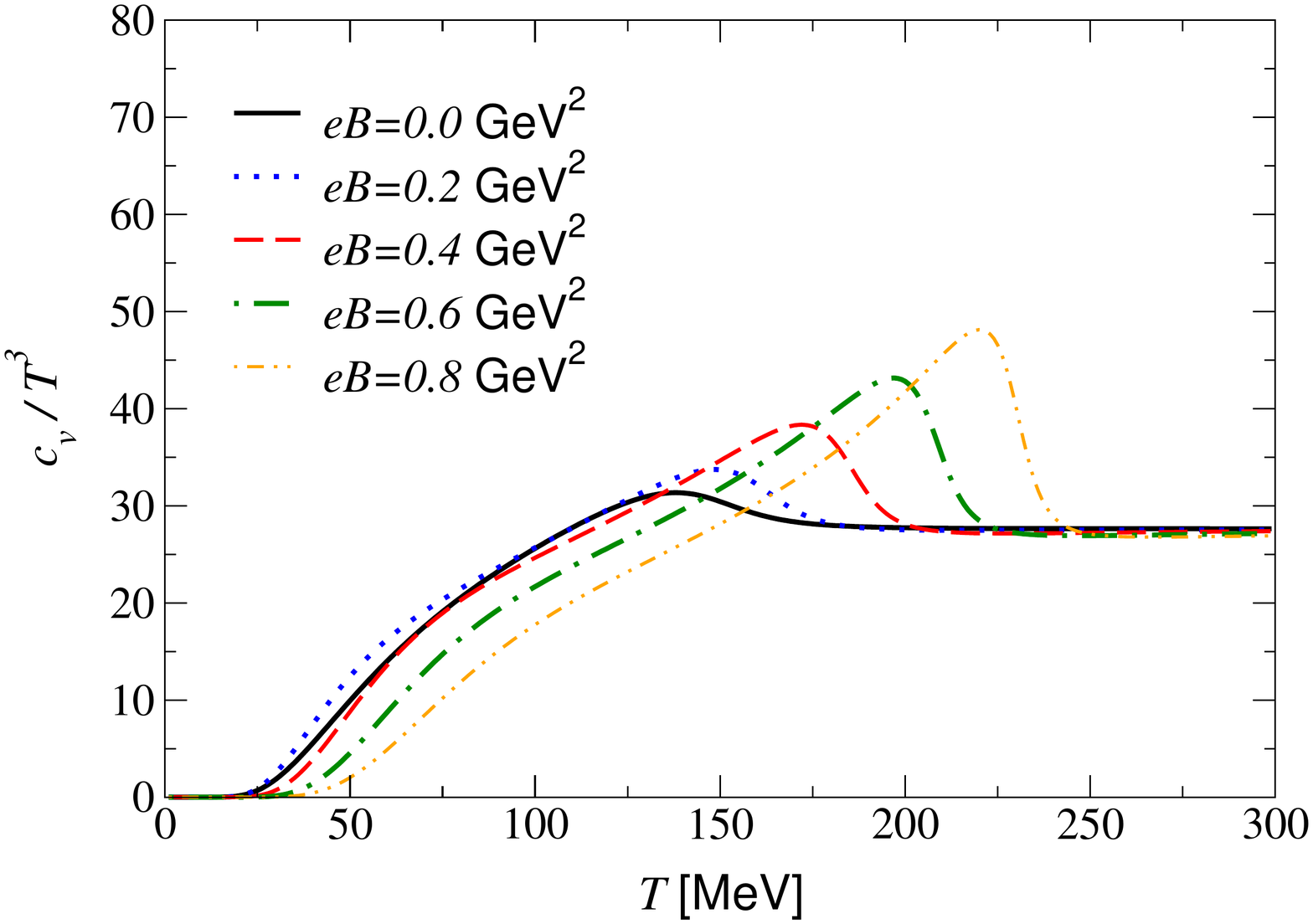}\hspace*{0.0cm}
\includegraphics[width=0.5\linewidth,angle=0]{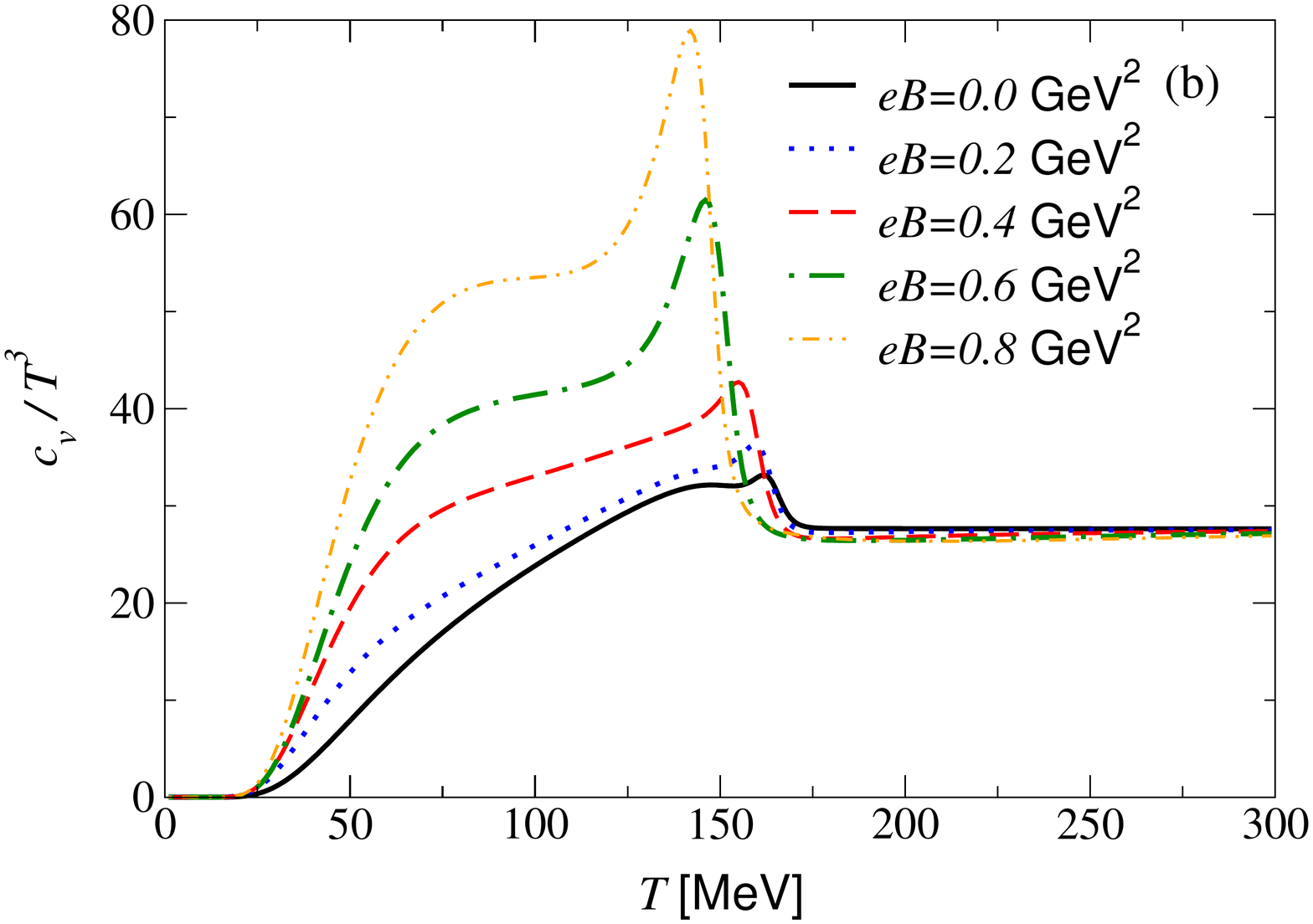}
\caption{The thermal susceptibility and specific heat as functions of temperature 
for different values of the magnetic field obtained with $G$ (left) and $G(B,T)$ (right).}
\label{fig5}
\end{center}
\end{figure*}

\begin{figure*}[hbt]
\begin{center}
\includegraphics[width=0.5\linewidth,angle=0]{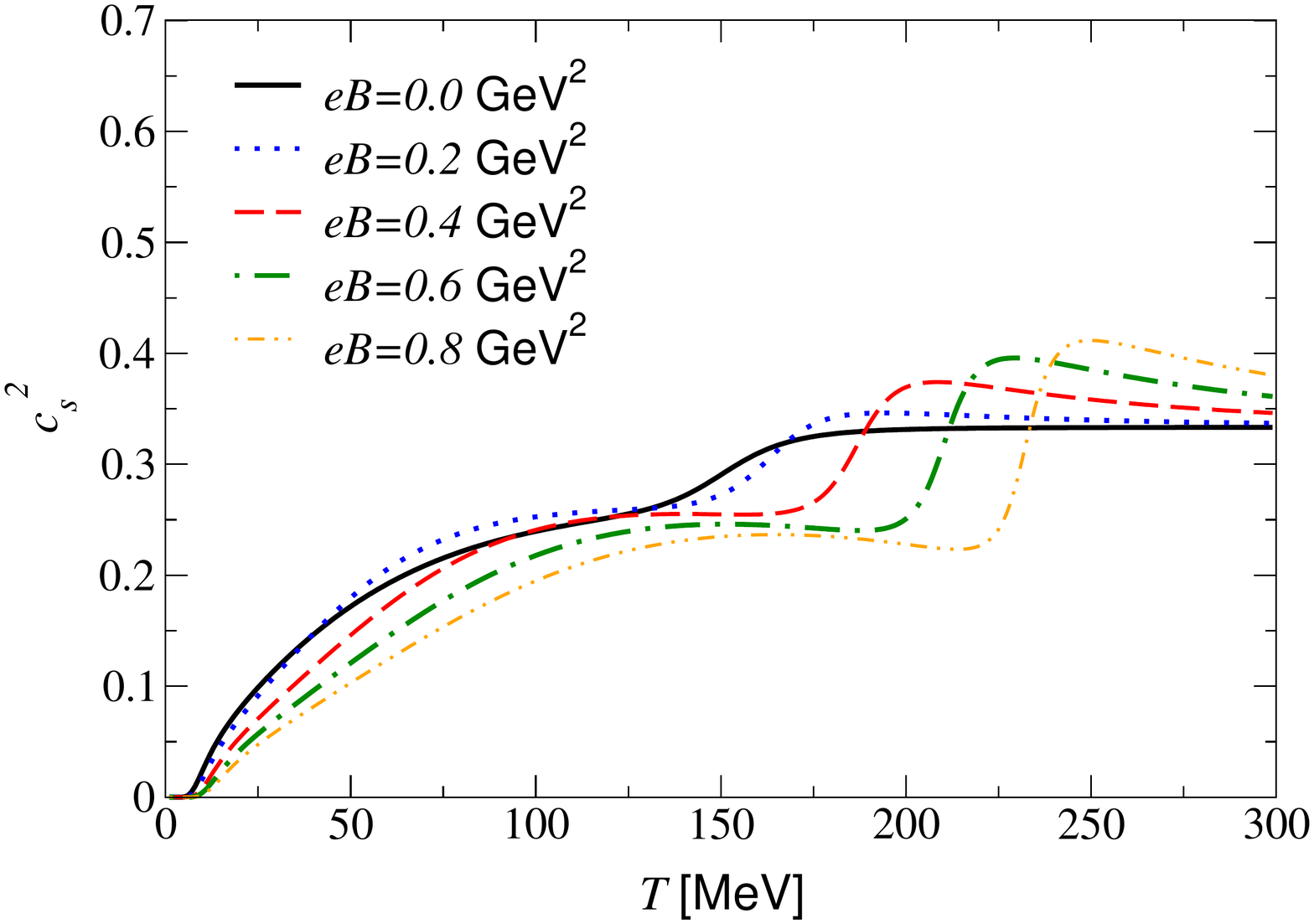}\hspace*{0.0cm}
\includegraphics[width=0.5\linewidth,angle=0]{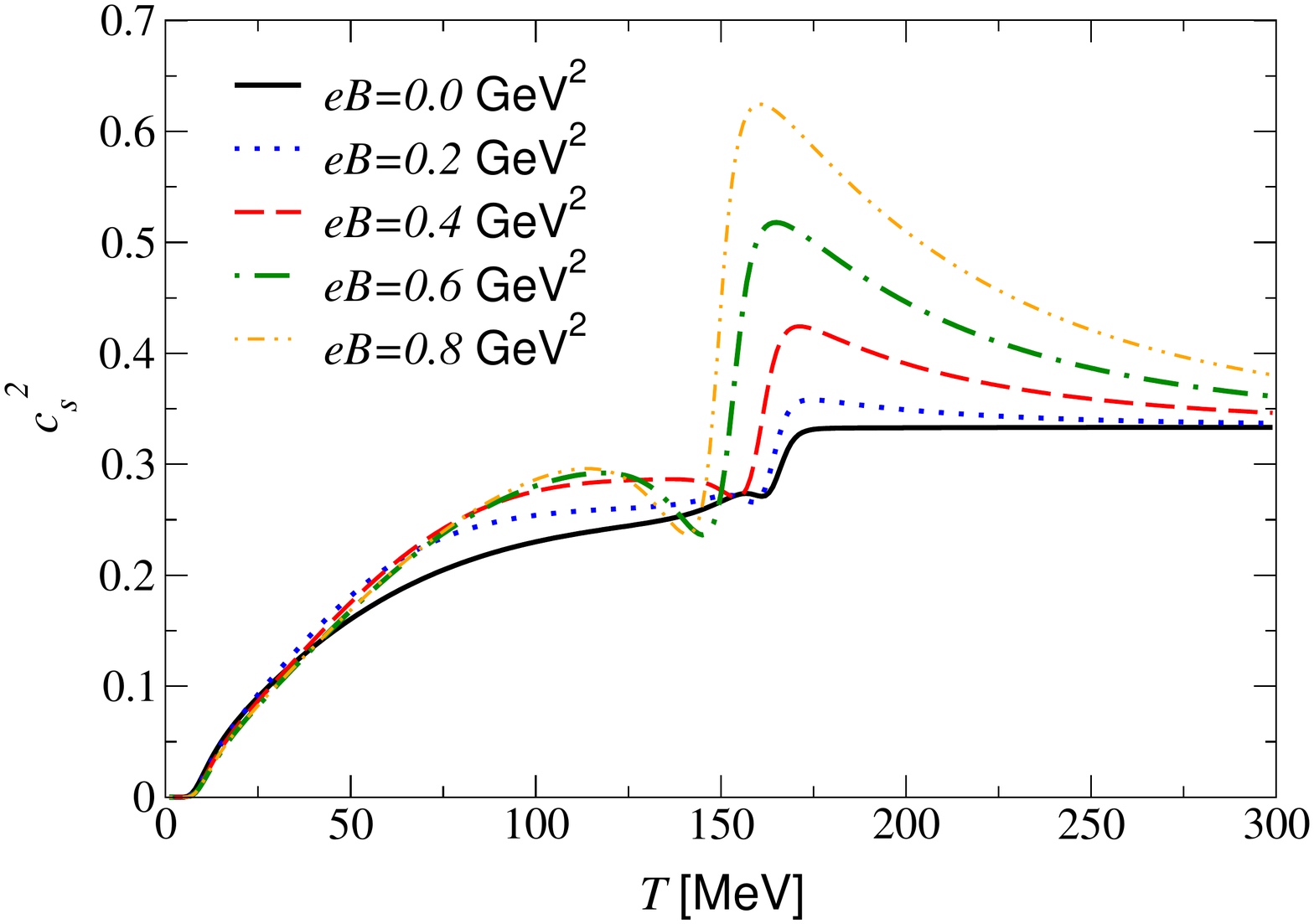}
\vspace*{0.0cm}
\includegraphics[width=0.5\linewidth,angle=0]{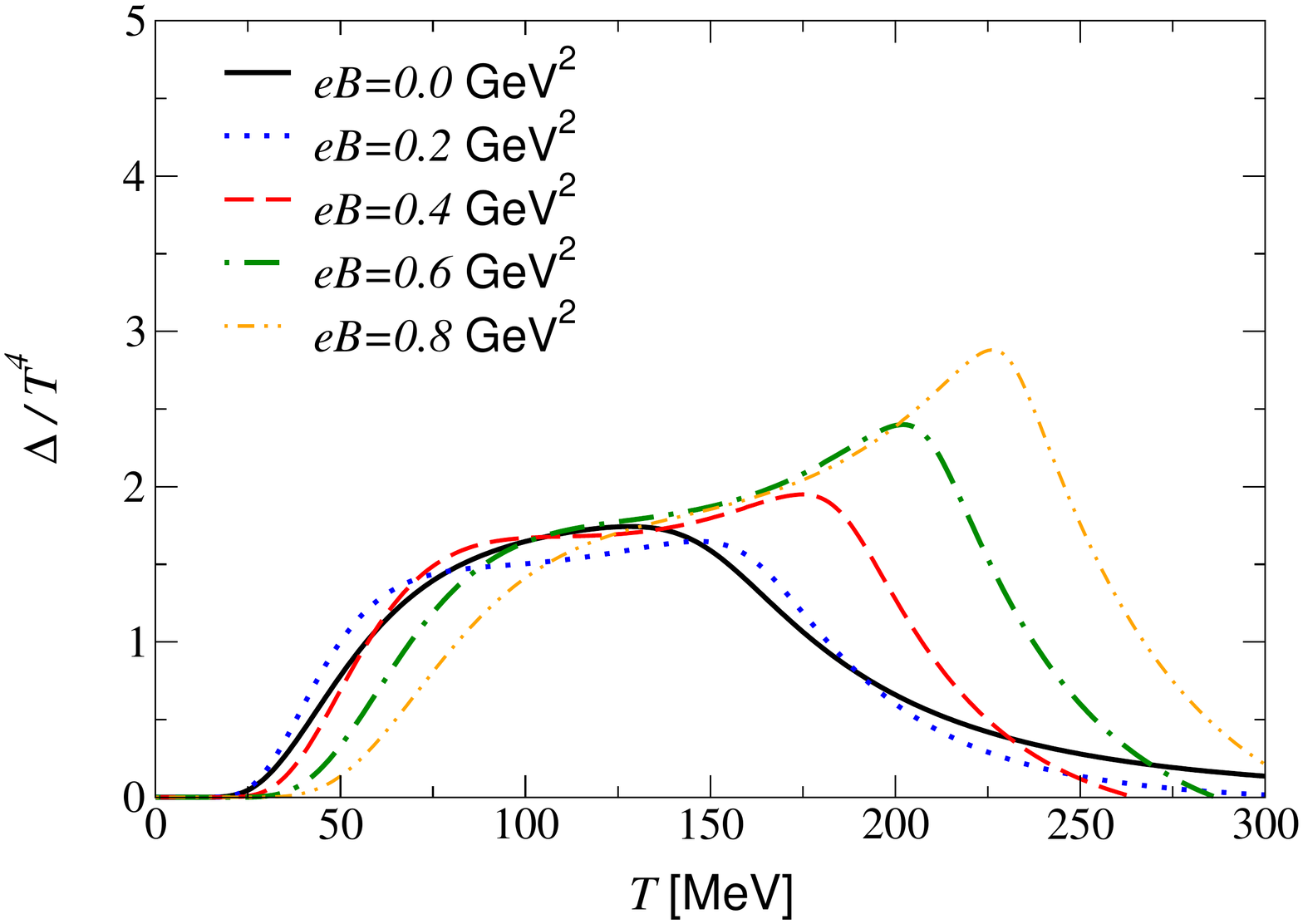}\hspace*{0.0cm}
\includegraphics[width=0.5\linewidth,angle=0]{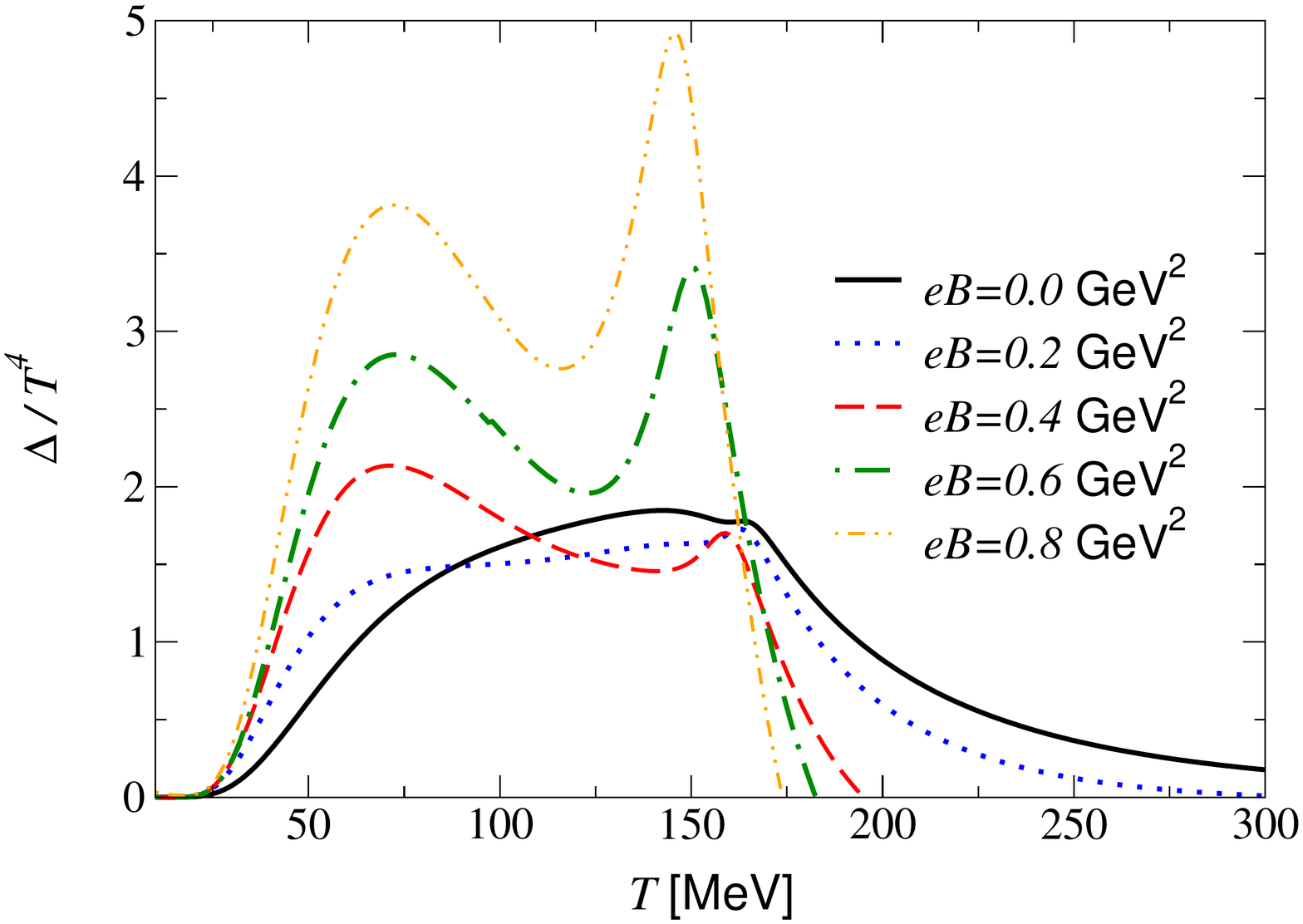}
\caption{The sound velocity squared and
interaction measure as functions of temperature for different values of the magnetic 
field obtained with $G$ (left) and $G(B,T)$ (right).}
\label{fig6}
\end{center}
\end{figure*}

Fig. \ref{fig5} displays $ \chi_T$ and $c_v$, and Fig.~\ref{fig6} displays $c_s^2$ and $\Delta$. 
As in the previous cases, we  observe an overall enhancement of all quantities in the transition
region for strong magnetic fields while the Stefan-Boltzmann limit is approached as the temperature
increases. 

The results clearly indicate that the thermal susceptibility changes dramatically when 
$G$ is replaced by $G(B,T)$. In particular, one notices in Fig.~\ref{fig6} that $\Delta$ presents
peaks that move in the direction of low temperatures when $B$ increases, in accordance with 
Ref.~\cite {bali}. Remark that the maxima that appear at low (and intermediate) values of $T$ are 
due the parametrization of the model, because the value of the coupling that fits the lattice 
results at $T=B=0$ is $10\,\%$ smaller than the usual value for the model \cite{jpcsGB0}. As we have verified, 
using this value of $G$ that fits the lattice results, the shape for $\Delta$ is deformed for 
small (intermediate) values of $T$, but since this region has not yet been explored by LQCD 
simulations, we cannot draw further conclusions at this point.  
At this point it is important to emphasize that the critical temperature for vanishing $B$ defined  as the value, $T_{pc}$, where the thermal susceptibility reaches a peak does not coincide with the critical temperature found by lattice simulations. This discrepancy shows up for instance on the top right panel of Fig.\ref{fig5} for the case of $B=0$. 
Effective models do not reproduce the same lattice values for $T_{pc}$ at $B=0$ and, usually, most authors linearly rescale the temperature axis so that their  results  match the lattice inflection point at $B = 0$. In our case we do not rescale the temperature axis, the lattice results are given at $B=T=0$. We perform our fitting at the critical region extrapolating the results for $T=0$ as well as  high-$T$ so that our predictions  for $T_{pc}$, at $B=0$, are slightly different from those obtained by LQCD. 

\begin{figure*}[h]
\begin{center}
\includegraphics[width=0.4\linewidth,angle=0]{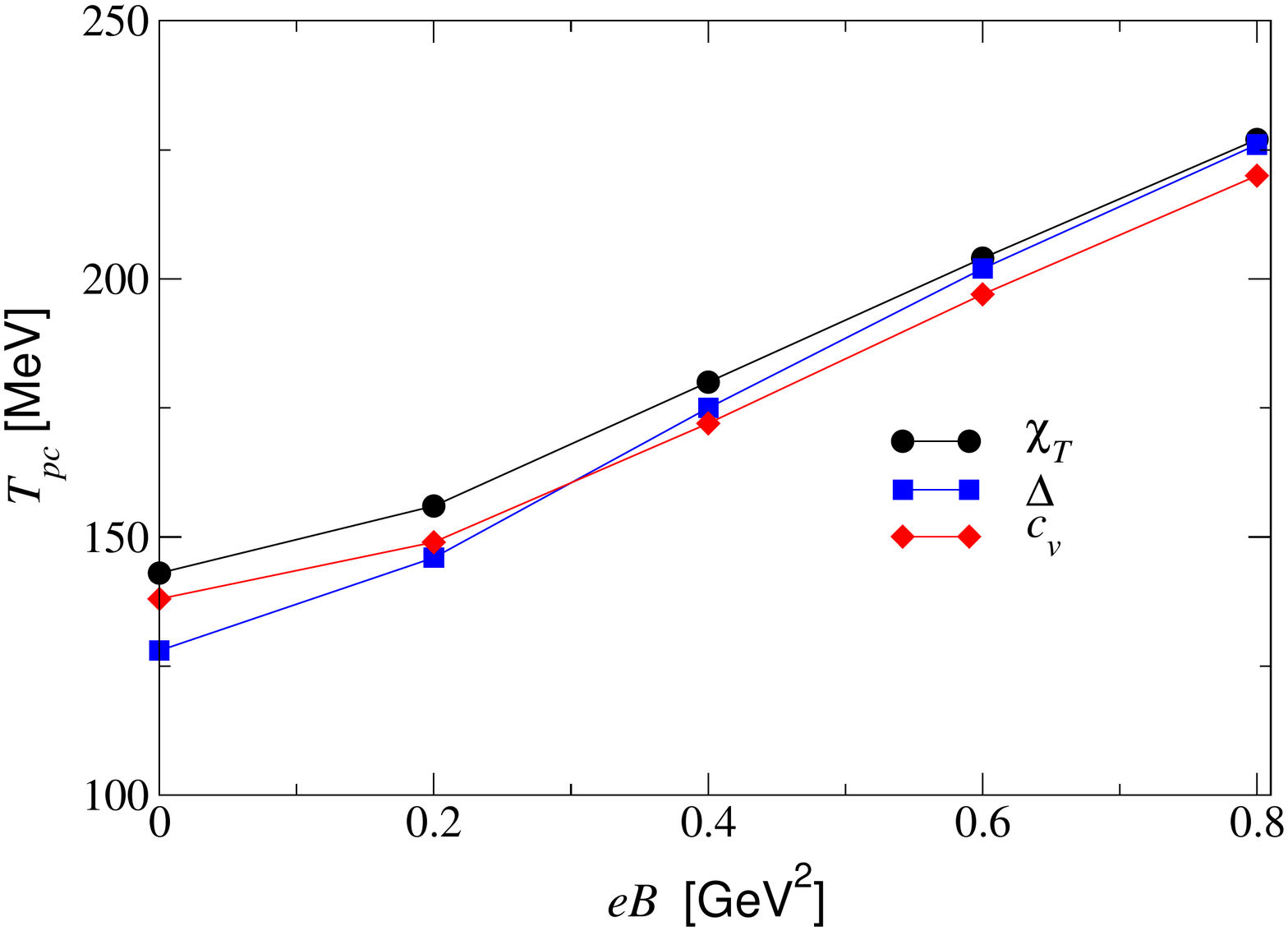} 
\includegraphics[width=0.4\linewidth,angle=0]{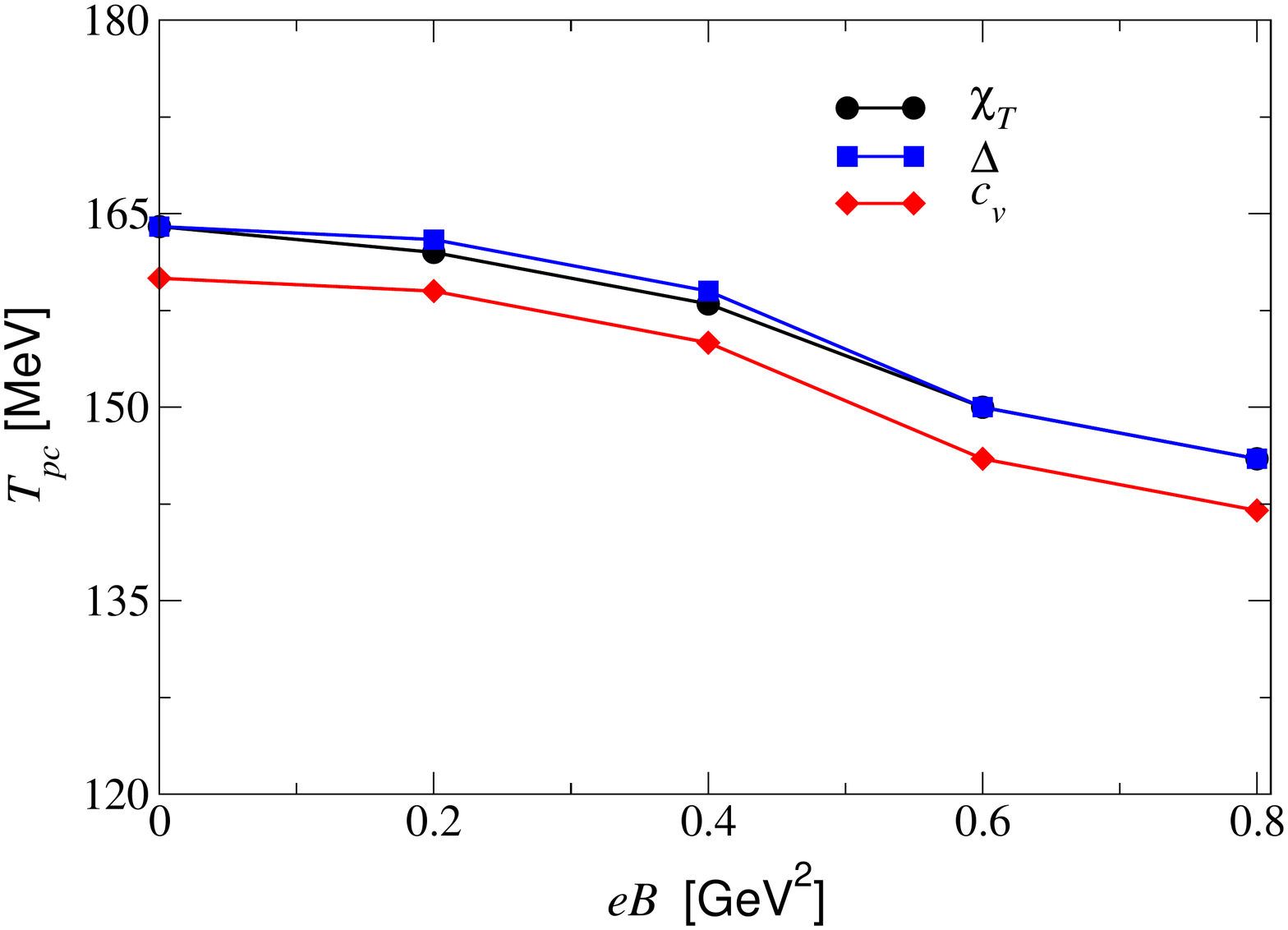}
\caption{The pseudocritical temperature for the chiral transition of magnetized 
quark matter as a function of the magnetic field strength obtained with $G$ (left) 
and with $G(B,T)$ (right).}
\label{fig7}
\end{center}
\end{figure*}

The dependence of the pseudocritical temperature on the magnetic field 
strength is displayed in Fig.~\ref{fig7}, which shows that when $G(B,T)$ is used, 
the phenomenon of IMC is observed to occur in a manner consistent with LQCD predictions.  
In this figure we define $T_{pc}$ using the thermal susceptibility $\chi_T$ and the 
specific heat $c_v$; we also include the temperatures of the maxima of interaction 
measure $\Delta$ to investigate its displacement with increasing $B$. It is interesting 
to remark that although this peak appears at a temperature which is a little higher 
than $T_{pc}$, it approximately follows the behavior of magnetic thermal susceptibility.

Finally, let us consider the magnetization which, in our case, can be written as

\begin{equation}
{\cal M}=\frac{d P}{dB} = \frac{\partial P}{\partial B}
+ \frac{\partial P}{\partial M}\frac{\partial M}{\partial B}
+ \frac{\partial P}{\partial G}
\frac{\partial G}{\partial B} ,
\label{mag-def}
\end{equation}
but, the quark masses are obtained by the gap equation $\partial P/{\partial M} = 0$, so that the second term vanishes. 
Notice that a linear term, arising from the $B^2/2$ contribution to the pressure, has 
been neglected so as to normalize $\cal M$ to vanish at zero temperature. 
Therefore, 
\begin{eqnarray}
{\cal M} &=&  \frac{\partial}{\partial B} \left(P_u + P_d\right) 
+ \frac{(M-m)^2}{4G^2} \frac{\partial G}{\partial B} . 
\label{eq:mag}
\end{eqnarray}

\noindent
Since the vacuum part of the pressure do not depend on $B$, 
they do not contribute to the magnetization. The derivatives of the 
pressure are

\begin{eqnarray}
\frac{\partial P^{mag}_f}{\partial B}  &=& \frac{2 P^{mag}_f}{B} -
\frac{N_c |q_f| }{4\pi ^2} M^2  \biggl[ \ln
\Gamma(x_f) - \frac{1}{2} \ln (2 \pi) \nonumber\\
&+&  x_f - \left(x_f - \frac{1}{2}\right) \ln(x_f) \biggr] \,,
\label{derPmag}
\end{eqnarray}

\begin{eqnarray}
\frac{\partial P^{Tmag}_f}{\partial B} &=& \frac{P_f^{Tmag}}{B}\nonumber\\
&& \, - \frac{N_c|q_f|^ 2 B}{2\pi^2} \sum_{k=0}^{\infty} k \alpha_k  
\int_{-\infty}^{+\infty} dp \frac{n(E_f)}{E_f}.
\label{derPTmag}
\end{eqnarray}
The magnetization, Eq.~(\ref{eq:mag}), is readily obtained from the expressions given in Sec.~II for the pressure. 
The remaining derivatives are easily calculated.

\begin{figure*}[h]
\begin{center}
\includegraphics[width=0.4\linewidth,angle=0]{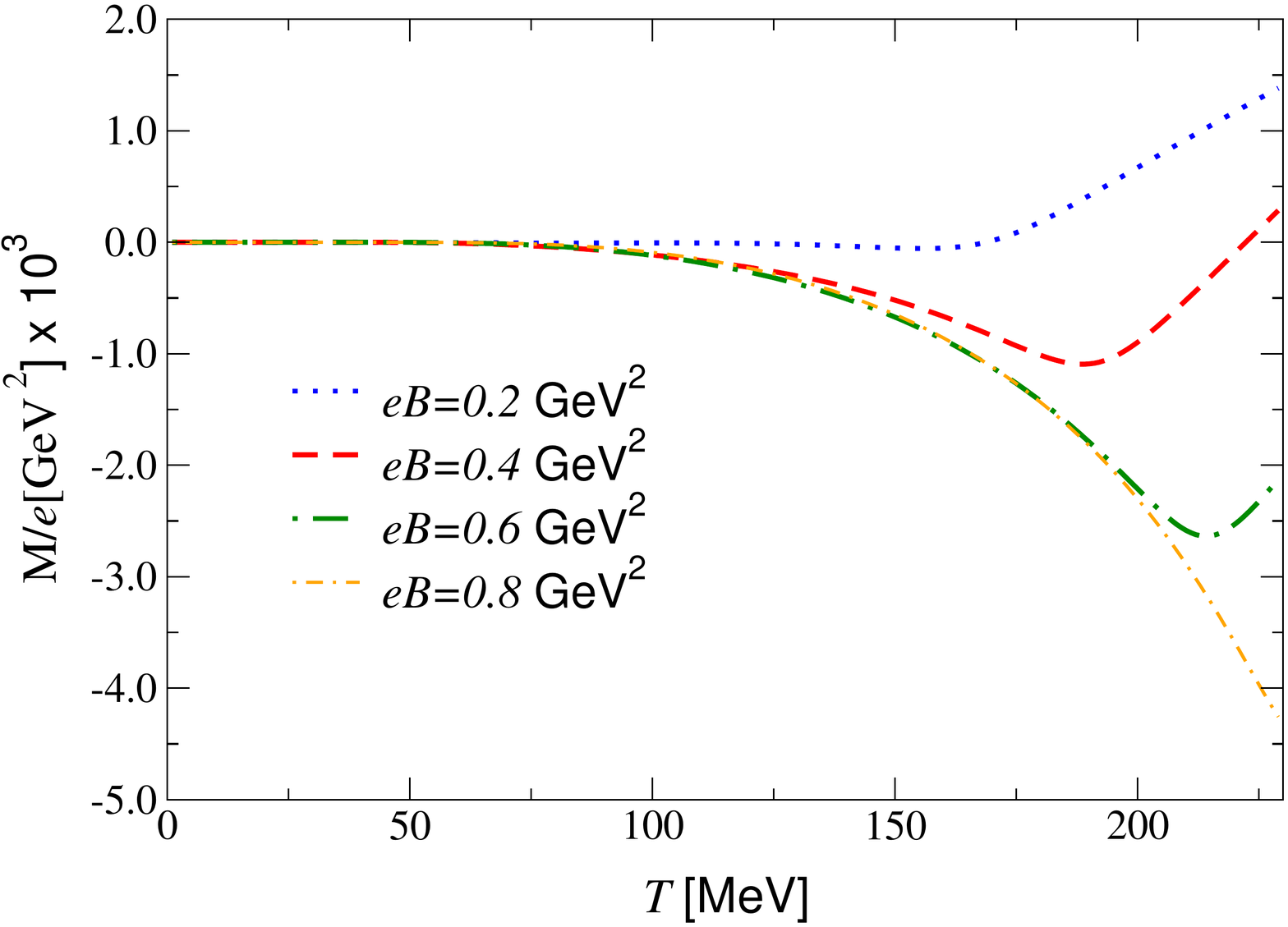} 
\includegraphics[width=0.4\linewidth,angle=0]{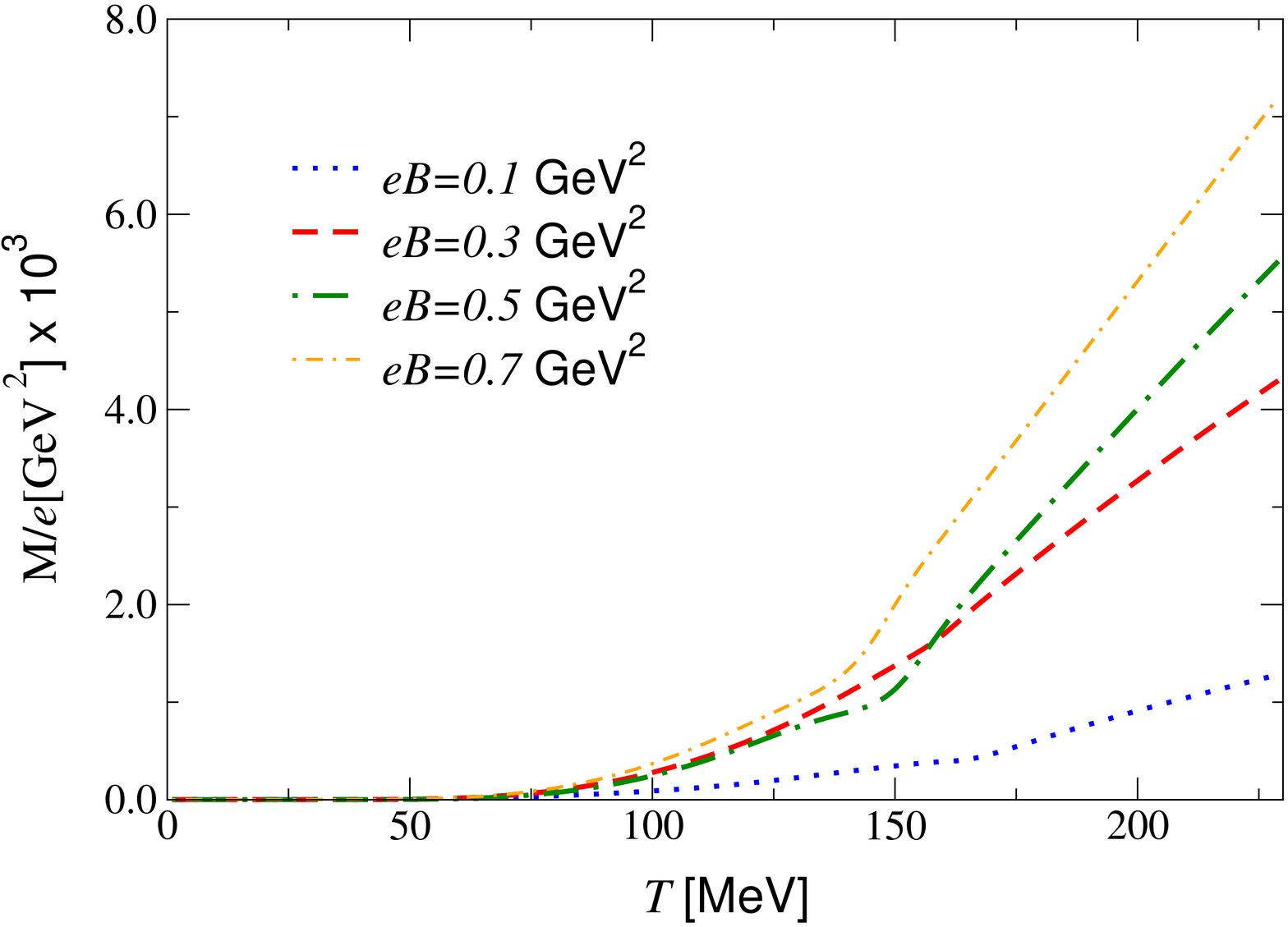}
\caption{The normalized magnetization $M/e$ of quark matter as a function of the temperature 
for different values of the magnetic field strength obtained with $G$ (left) and 
$G(B,T)$ (right).}
\label{fig8}
\end{center}
\end{figure*}

\begin{figure*}[h]
\begin{center}
\includegraphics[width=0.4\linewidth,angle=0]{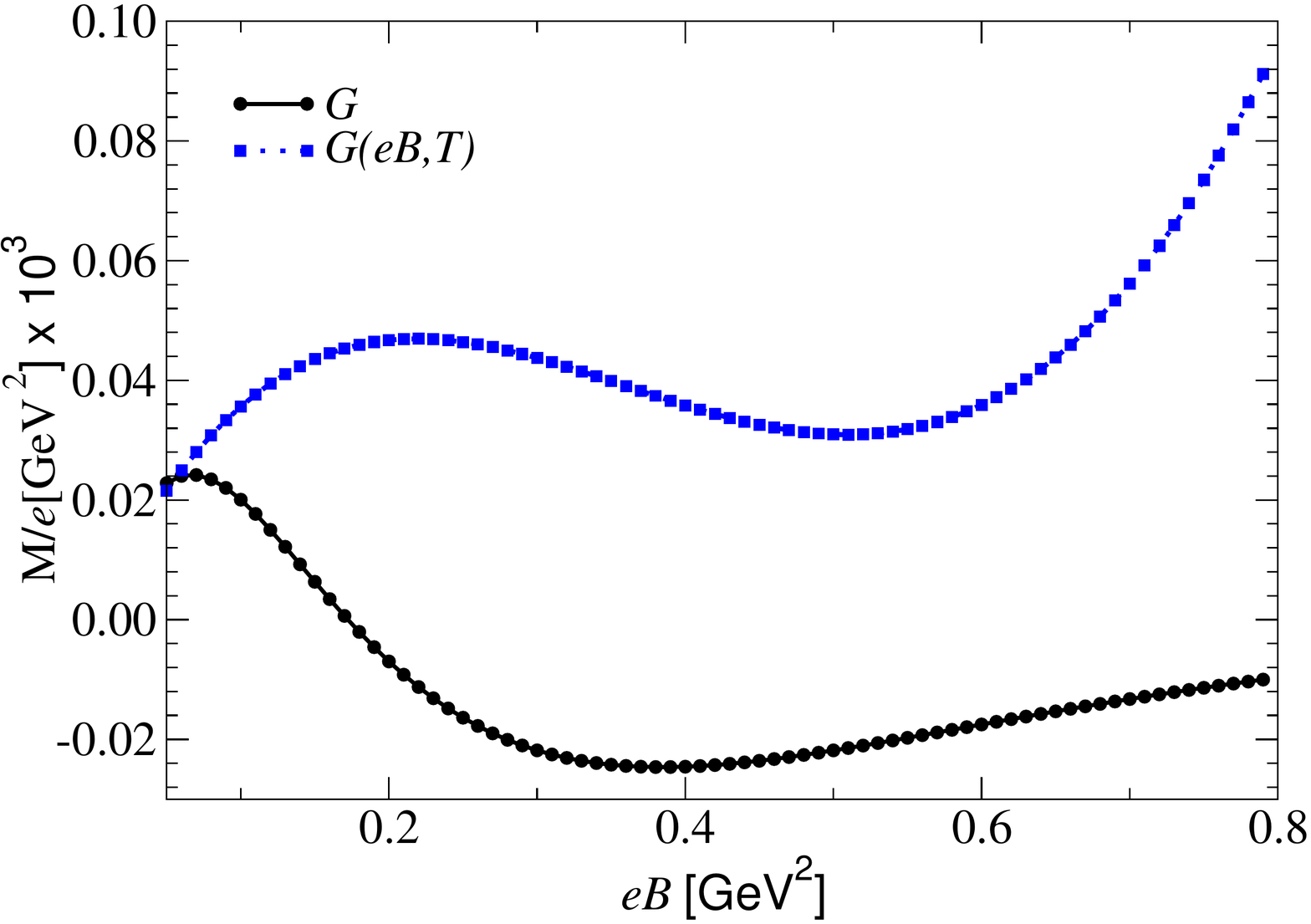}\hspace*{0.0cm}
\includegraphics[width=0.4\linewidth,angle=0]{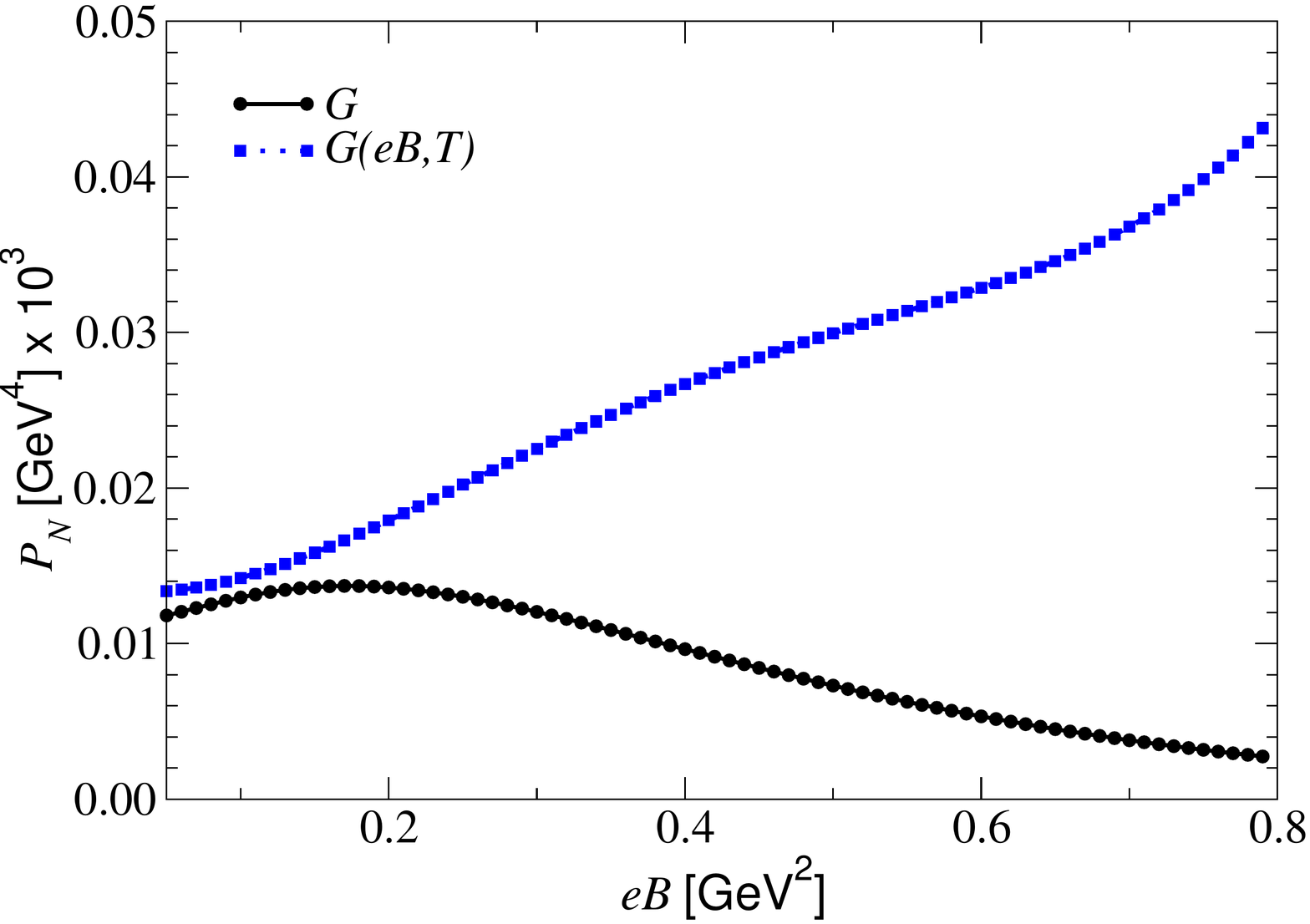}
\caption{The normalized magnetization $M/e$ (left) and the normalized pressure (right) 
as a function of the magnetic field at $T = 70$ MeV, obtained with $G$ (solid line) and  
$G(B,T)$ (dotted line).}
\label{fig9}
\end{center}
\end{figure*}

In Fig. \ref {fig8} we show the normalized magnetization ${\cal M}/e$ as a function of 
temperature for different magnetic field strengths. Again, one observes that a fixed coupling 
$G$ does not predict a monotonic increase of the magnetization with $eB$ for a given temperature. 
This can be observed more clearly in Fig.~\ref{fig9} where we show how the pressure and magnetization 
depend on $eB$ at a fixed temperature, $T=70$~MeV. While the traditional fixed coupling predicts a 
magnetization that becomes negative as the magnetic field strength is increased, the thermo-magnetic 
coupling yields to positive magnetizations and is in agreement to the paramagnetic nature of thermal QCD 
medium observed in $N_f=2+1$ LQCD simulations of Ref.~\cite{bali}.


We close this section by remarking that, to the best of our knowledge, LQCD predictions for 
the $N_f=2$ case analyzed here are not available in the literature.  Although one could argue 
that the use of a three flavor version of the NJL model would be more appropriate to compare with
lattice results, we recall that our ansatz for the four-fermion interaction strength, $G$, was 
obtained by fitting the LQCD results for the light quark sector, which represents the relevant 
degrees of freedom regarding the chiral transition. Using this fit, one retrieves, at least 
qualitatively, most of the lattice predictions for different thermodynamical quantities for the $N_f=2+1$
case, improving over predictions made with a fixed coupling. Remark that a more sophisticated $SU(3)$ 
NJL model possesses a six-fermion vertex characterized by another coupling, $K$, tailored to account 
for strangeness, and has been adopted for realistic astrophysical applications, where strangeness is 
important, or comparisons aiming at quantitative agreement with $N_f=2+1$ LQCD predictions 
for thermodynamical observables. In principle, $K$ can also be considered to have a thermo-magnetic 
dependence and then with this extra degree of freedom one could attempt to obtain a numerically more
accurate description of the LQCD results for $N_f=2+1$ as a (more appropriate) alternative to the simple approach
considering solely the $G$ coupling, with a magnetic dependence only~\cite{prdferreira}.
In a forthcoming work, we will show the thermo-magnetic dependencies $G(B,T)$ and $K(B,T)$ obtained 
by fitting the mean and difference of $u$ and $d$ quark condensates computed in the framework of LQCD. 
While finishing our paper we have learned of a similar implementation of $G(B,T)$ in Ref.~\cite{ayalafit}

%%%%%%%%%%%%%%%%%%%%%%%%%%%%%%%%%%%%%%%%%%%%%%%%
\section{Conclusions}

We have investigated the  thermodynamics of magnetized quark matter within the NJL 
model using a coupling $G$ that decreases with both the temperature $T$ and the magnetic field 
$B$. The $T$ and $B$ dependence of $G$ was obtained by an accurate fit of lattice QCD
results for the light-quark condensates. Using the fitted $G(B,T)$, we computed different 
thermodynamical quantities and analyzed the qualitative changes implied by the fitted coupling. 
The main conclusion of our work is that a coupling $G(B,T)$ that fits lattice result for $T_{pc}$ as 
determined by the quark condensates, gives results for the pressure, entropy and energy densities 
that are in qualitative agreement with corresponding lattice results, while a $B-$ and 
$T-$in\-de\-pend\-ent coupling gives qualitatively different results for those quantities. 
In particular, for any fixed temperature, quantities such as pressure and magnetization 
obtained with $G(eB,T)$ increase with $eB$, a result that is consistent with lattice QCD results.

Here, we have shown a very important result: NJL model calculations performed with 
our thermo-magnetic coupling predicts that the magnetization is positive in all temperature range, 
which complies to the paramagnetic nature of QCD medium and is in agreement with lattice calculations. 
Another feature that supports the thermo-magnetic dependence of the coupling constant is the observation 
that the chiral transition seems sharper and peaks observed in quantities such as the entropy density increase 
considerably with $eB$, a feature that is also consistent with lattice simulations and often
missed when using a $B-$ and $T-$independent coupling. As remarked earlier, the results seem to 
indicate that the $B$ and $T$ dependence in $G$ that gives the correct $T_{pc}$ is neither fortuitous nor 
valid for a single physical quantity only; it seems to capture correctly the physics left out in 
the conventional NJL model. Also, any interpolation formula of the lattice data points for the 
quark condensate is expected to give qualitatively similar results for the  thermodynamical quantities 
in the appropriate range of $T$ and $B$.

Our results indicate that it is crucial to take into account both $B$ and $T$ effects in
the effective coupling. First, it is virtually impossible to fit lattice results with an
effective NJL coupling that depends on $B$; a coupling that depends on $B$, despite decreasing
with $B$, leads~\cite{prdferreira} to non monotonic decrease of $T_{pc}$ at $eB \approx 1.1 \, 
{\rm GeV}^2$. It is well known that in QCD the temperature also represents an energy scale and the 
coupling constant runs with $T$ when $B=0$. Therefore, in principle, purely thermal effects should also 
influence the NJL coupling -- and of course, the same is expected for six-fermion or higher-order 
NJL couplings. However, in practice, with few exceptions~\cite{bernard,marcus}, 
purely thermal effects are usually neglected since no qualitative discrepancies between lattice and 
model predictions have been observed so far when $T\ne 0$ and $B=0$, in contrast to the case when 
$B \neq 0$. Indeed, our results show that to have a consistent monotonic decrease of $T_{pc}$ with 
$B$, it is crucial to consider a $B-T$ dependent coupling, which seems to be consistent with the 
findings of Ref.~\cite {bruno}, where the authors of that reference argue that chiral models with 
couplings depending solely on $B$ are unable to correctly describe IMC. Also,  in Ref.~\cite {mao}.
IMC is observed when a thermo-magnetic effective coupling appears as a consequence of improving 
on a mean field evaluation with mesonic effects. Obviously, the model itself cannot explain the basic
physics behind the required $B$ and $T$ dependence of the effective coupling, but it seems consistent with
a physical interpretation based on competing effects between quark and gluon charges, as demonstrated 
in the one-loop vertex correction calculated in Ref.~\cite{ayala7}. Naturally, other interpretations are
not excluded and further work is required to clarify the physical picture.

In summary, we have shown that the NJL model can be patched in order to accurately reproduce IMC 
which is observed to take place within the chiral transition of hot and magnetized quark matter. 
In particular, the thermo-magnetic dependent coupling Eq.~(\ref {ourGBT}) seems to provide an appropriate 
effective coupling $G(B,T)$ that captures effects beyond the conventional NJL models and which can be promptly 
employed to improve predictions to hadronic systems involving large magnetic fields. The apparent weak dependence 
on $eB$ is essential to obtain a positive magnetization while the sharp dependence on $T$ ensures a good description 
of the chiral phase transition.

%%%%%%%%%%%%%%%%%%%%%%%%%%%%%%%%%%%%%%%%%%%%
\clearpage
\vskip 0.2cm
\noindent
{\bf Acknowledgments}

\smallskip\noindent
We thank G. Endrodi for discussions and also for providing the lattice data of the up and down quark condensates, and A. Ayala 
for useful comments on an earlier version of the manuscript. M. B. P. is also grateful to E. S. Fraga for useful comments. 
This work was supported by CNPq grants 475110/2013-7, 232766/2014-2, 308828/2013-5 (R.L.S.F), 306195/2015-1 (VST), 307458\\/2013-0 (SSA), 303592/2013-3 (MBP), 305894/2009-9 (GK),  FAPESP grants 2013/01907-0 (GK), 2016/07061-3 (VST)
and FAEPEX grant 3284/16 (VST). R.L.S.F. acknowledges the kind hospitality of the Center for Nuclear Research at Kent State University, where part of this work has been done. 

%%%%%%%%%%%%%%%%%%%%%%%%%%%%%%%%%%%%%%%%%%%%%%%%%%%%%%%%%%%%%%%%%%%%%%%%%%%
%
%

\end{document}